\documentclass[prd,aps,tightenlines,twocolumn,nofootinbib,preprintnumbers,showkeys]{revtex4}
\pdfoutput=1
\usepackage{amsmath,amssymb}
\usepackage{epsfig}
\usepackage{graphicx}
\usepackage{subfigure}

\usepackage{xcolor}
\usepackage{soul}

\usepackage{slashed}

\newcommand{\be}{\begin{equation}}
\newcommand{\ee}{\end{equation}}
\newcommand{\bea}{\begin{eqnarray}}
\newcommand{\eea}{\end{eqnarray}}
\newcommand{\mg}{m_{3/2}}

\newcommand{\mpl}{M_{Pl}}
\newcommand{\half}{\frac{1}{2}}

\newcommand{\rad}{\sqrt{2}}
\newcommand{\cub}{\sqrt{3}}
\newcommand{\lsim}{\mbox{\raisebox{-.6ex}{~$\stackrel{<}{\sim}$~}}}
\newcommand{\gsim}{\mbox{\raisebox{-.6ex}{~$\stackrel{>}{\sim}$~}}}
\newcommand{\G}{\mathrm{GeV}}
\newcommand{\T}{\mathrm{TeV}}
\newcommand{\grav}{{3/2}}
\newcommand{\order}{\mathcal{O}}


\begin{document}

\title{Reheating, thermalization and non-thermal gravitino production in MSSM inflation}
\author{Andrea Ferrantelli}

\affiliation{
Tallinn University of Technology, Faculty of Civil Engineering, 19086 Tallinn, Estonia}

\date{\today}

\begin{abstract}

In the framework of MSSM inflation, matter and gravitino production are here investigated through the decay of the fields which are coupled to the \textit{udd} inflaton, a gauge invariant combination of squarks.
After the end of inflation, the flat direction oscillates about the minimum of its potential, losing at each oscillation about 56\% of its energy into bursts of gauge/gaugino and scalar quanta when crossing the origin. These particles then acquire a large inflaton VEV-induced mass and decay perturbatively into the MSSM quanta and gravitinos, transferring the inflaton energy very efficiently via instant preheating.

Regarding thermalization, we show that the MSSM degrees of freedom thermalize very quickly, yet not immediately by virtue of the large vacuum expectation value of the inflaton, which breaks the $SU(3)_C\times U(1)_Y$ symmetry into a residual $U(1)$. The energy transfer to the MSSM quanta is very efficient, since full thermalization is achieved after only $\mathcal{O}(40)$ complete oscillations. The \textit{udd} inflaton thus provides an extremely efficient reheating of the Universe, with a temperature $T_{reh}=\mathcal{O}(10^8\mathrm{GeV})$ that allows for instance several mechanisms of baryogenesis.

We also compute the gravitino number density from the perturbative decay of the flat direction and of the SUSY multiplet. We find that the gravitinos are produced in negligible amount and satisfy cosmological bounds such as the Big Bang Nucleosynthesis (BBN) and Dark Matter (DM) constraints.

\end{abstract}

\maketitle


\section{Introduction}\label{sec:intro}

The Minimal Supersymmetric Standard Model (MSSM) \cite{Martin:1997ns,Haber:1984rc} is believed to be a good candidate for an extension of the extremely successful Standard Model (SM) of Particle Physics. The MSSM scalar potential has many $D$-flat directions, which are classified by gauge-invariant monomials of the theory. Most of these carry baryon and/or lepton number \cite{Dine:2003ax,Gherghetta:1995dv}, and are relevant for the early universe cosmology \cite{Enqvist:2003gh}.

Among the number of models of primordial inflation which address the origin and properties of the inflaton field \cite{Mazumdar:2010sa,Allahverdi:2010xz}, MSSM inflation \cite{Allahverdi:2006iq,Allahverdi:2006we} assumes this particle to be no longer an unknown Standard Model (SM) gauge singlet, but instead one of these $D$-flat directions (f.d.). In particular, the inflaton can be a combination of either sleptons (the \textit{LLe} f.d.) or squarks (the \textit{udd}) \cite{Enqvist:2003gh,Dine:2003ax,Gherghetta:1995dv}.

In this paper we assume that the inflaton field is the \textit{udd} flat direction, a combination of squarks.
After the end of inflation, when its mass $m_\phi \sim H$, with $H$ the Hubble expansion rate, the \textit{udd} starts oscillating coherently about the minimum of its potential. At each origin crossing, it releases bursts of squarks, gluons and gluinos through a non-perturbative mechanism of preheating \cite{Felder:1998vq,Allahverdi:2000fz}.

The condensate VEV, which we call $\phi$, spontaneously breaks the original $SU(3)_C\times U(1)_Y$ SM gauge symmetry to a 
$U(1)$, which is a linear combination of $I_3$ and $U(1)_Y$.
As we will show, this induces a large mass for the squarks and gluon/gluino fields $m_{\mathrm{eff}}\propto g\phi$ via the Higgs mechanism ($g$ is a coupling). These will then decay perturbatively into the MSSM degrees of freedom: at each oscillation, nearly $56\%$ of the inflaton energy is transferred to the observable sector via \textit{instant preheating} \cite{Traschen:1990sw,Shtanov:1994ce,Kofman:1994rk,Kofman:1997yn,Allahverdi:2008pf}. 

The thermalization of SM quanta is not instantaneous by virtue of the above gauge symmetry breaking; however, reheating in MSSM inflation is generally very efficient \cite{Allahverdi:2006we,Allahverdi:2011aj}. The Universe thermalizes within one Hubble time after the end of inflation, since the frequency of oscillations is $\sim 10^3\, H$. Specifically, here we show that the \textit{udd} inflaton provides full thermalization after only $\mathcal{O}(40)$ oscillations, with a temperature $\mathcal{O}(10^{8}\,\G)$. This is perhaps the quickest and most efficient way to thermalize the Universe, even within MSSM inflation \cite{Allahverdi:2011aj}.


We also compute for the first time gravitino production from perturbative decay processes of the \textit{udd} inflaton and of the fields produced during its oscillations, investigating whether there might be a gravitino problem \cite{Khlopov:1984pf,Moroi:1995fs,Kawasaki:2006hm}. Consistently with the literature, we verify that the number density of gravitinos produced by the inflaton is not problematic \cite{Allahverdi:2000fz,Allahverdi:2008pf}, since both cosmological constraints on Big Bang Nucleosynthesis (BBN) and the observed Dark Matter (DM) abundance are satisfied.

  The same happens for the decay of particles produced by the flat direction at each origin crossing, regardless of their large VEV-induced effective mass (of the order $\sim 10^{15}\G$) obtained when the inflaton rolls back its potential, away from the minimum. We obtain a yield value that is well below dangerous values for any realistic gravitino mass.


The present article is organized as follows: in Section \ref{sec:flatdirection} we introduce the \textit{udd} inflaton and describe its structure, potential, dynamics and energy scales. In Section \ref{sec:couplings} we compute the couplings of the SUSY multiplet to this flat direction and their decay widths into the MSSM degrees of freedom.
In Section \ref{sec:reheating-after-mssm-inflation} we study the reheating and thermalization process, while in Section \ref{sec:non-thermal-gravitino-production-in-mssm-inflation} we investigate the non-thermal gravitino production from the perturbative decay of the flat direction and of the fields to which it is coupled. After our Conclusions, in Appendix \ref{sec:essential-expressions-for-the-bf-udd-inflaton} we list all the interactions of the SUSY multiplet with the MSSM.

\section{The udd flat direction}\label{sec:flatdirection}

The flat direction discussed in this paper is comprised of squarks with a family combination; $\mathbf{d_2d_3u_1}$. The three superfields are in triplet representations of $SU(3)_C$, which are generated by $T^a=(1/2)\lambda^a$, where $\lambda^a$ are the eight Gell-Mann matrices ($a=1,...,8$).
We identify the corresponding scalars with the right-handed squarks $\tilde{u}_R$ and $\tilde{d}_R$ of the MSSM, which are $SU(2)_W$ gauge singlets.
The quantum numbers are respectively $C=1,1,1$ and $Y=2/3,2/3,-4/3$, the $U(1)_Y$ hypercharge. The flat direction is then parametrized by the following~\cite{Allahverdi:2006iq,Allahverdi:2006we,Allahverdi:2008pf},
\be
\tilde{d}_2=\frac{\varphi}{\sqrt{3}}\left( \begin{array}{c}
1 \\
0 \\
0 \end{array} \right),\;
\tilde{d}_3=\frac{\varphi}{\sqrt{3}}\left( \begin{array}{c}
0 \\
1 \\
0 \end{array} \right),\;
\tilde{u}_1=\frac{\varphi}{\sqrt{3}}\left( \begin{array}{c}
0 \\
0 \\
1 \end{array} \right),
\ee
where $\tilde{d}_2$, $\tilde{d}_3$, $\tilde{u}_1$ are the scalar components of the corresponding superfields. The inflaton will be the real part of $\varphi$: $\phi = \varphi_R$.
We now summarize for completeness some general features of inflection point inflation \cite{Enqvist:2010vd,Allahverdi:2011aj}. 

During inflation, the inflaton $\phi$ acquires a large VEV resulting from the accumulation of quantum fluctuations\footnote{In this paper $\phi$ labels also the VEV, with abuse of notation.}. This leads to the formation of
a condensate along the flat direction \cite{Enqvist:2003gh}.
The potential along a complex flat direction $\phi$ can be written generally as (see e.g. \cite{Allahverdi:2008pf}),
\be \label{genflatpot}
V(\phi) = m^2 |\phi|^2 + \lambda^2 \frac{|\phi|^{2(n-1)}}{\mpl^{2(n-3)}} 
 +\left( A \lambda \frac{\phi^n}{\mpl^{n-3}} + \mathrm{h.c.}\right)
 \, ,
\ee
where $m, A\sim{\cal O}(\T)$ are respectively the soft supersymmetry breaking mass and the $A$-term. $\mpl=2.44\times 10^{18}\,\G$ is the reduced Planck mass, $n>4$ and usually $\lambda \sim  {\cal O}(1)$. More in general, instead of $\mpl$ any generic cutoff energy scale $M$ can be used in (\ref{genflatpot}) \cite{Dine:1995uk,Dine:1995kz}.

If we now assume $n=6$ and a real field, we obtain \cite{Allahverdi:2006iq},
\begin{equation} \label{flatpot}
V(\phi) = \half m^2_\phi \phi^2 - A \lambda \frac{\phi^6}{3 \mpl^{3}} + \lambda^2 \frac{{\phi}^{10}}{\mpl^{6}} \,,
\end{equation}
which has an inflection point for $A \approx \sqrt{40} m_{\phi}$ at  \cite{Enqvist:2003gh}
\begin{equation} \label{saddle}
\phi_0 = \left(\frac{m_\phi \mpl^{3}}{\sqrt{10} \lambda}\right)^{1/4}\,.
\end{equation}
For inflaton field values close to $\phi_0$, successful inflation can occur in the interval $\vert \phi - \phi_0 \vert \sim (\phi^3_0 / 60 \mpl^2)$.

Weak scale supersymmetry gives to the inflaton a mass $m_{\phi} \sim {\cal O}({\rm TeV})$, and the field VEV at their onset is $\phi_0 \sim {\cal O}(10^{14}\div10^{15})\,\G$.
	%
	Within this range of values, the \textit{udd} inflaton satisfies various cosmological bounds; as it was shown in \cite{Wang:2013qti}, the corresponding central value of the amplitude of seed perturbations $P_\xi$ and the spectral index $n_s$ are consistent with the Planck observations \cite{Ade:2013zuv}.

After inflation, the condensate VEV slowly rolls down its potential until the time when the Hubble expansion rate
$H(t)\sim m_\phi$, and the condensate VEV is $\phi_0\sim (m_\phi\mpl^{(n-3)})^{1/(n-2)}\gg m_\phi$ \cite{Dine:1995uk,Dine:1995kz}. At this point, it starts oscillating about $\phi = 0$ with a frequency $m_\phi$.

The Hubble rate during inflation is \cite{Enqvist:2010vd}
\be
H_{\rm inf}\sim \dfrac{V_0^{1/2}}{\cub\mpl}\propto \dfrac{m_\phi\phi_0}{\cub\mpl}\sim{\cal O}(\G)\,,
\ee
so the hierarchy $m_\phi \sim 10^3 H_{\rm inf}$ implies that, after inflation, the number of inflaton oscillations within a single Hubble time can be large. As we will show in Section \ref{sec:reheating-after-mssm-inflation}, this is important for reheating.

\section{Couplings of the inflaton and decay widths}\label{sec:couplings}

In this section we compute the decay widths of the fields which are coupled to the inflaton, namely the scalars, gauge bosons and gauginos, by using the couplings given in Appendix \ref{sec:essential-expressions-for-the-bf-udd-inflaton}. This supersymmetric multiplet belongs to representations of the gauge group $SU(3)_C\times U(1)_Y$, whose symmetry is broken to a residual $U(1)$ by the inflaton VEV.

\subsection{Scalar interactions and decay widths}

Since the Yukawa couplings are subdominant, we focus on the gauge interactions arising from $D$-terms. For $SU(3)_C$ we get \cite{Haber:1984rc,Martin:1997ns}:
\be\label{udd-scal-0}
D^a_{C}=-g_s\left(\tilde{u}^\dagger_1 T^a\tilde{u}_1+\tilde{d}^\dagger_2 T^a\tilde{d}_2+\tilde{d}^\dagger_3 T^a\tilde{d}_3\right)\,.
\ee
On the other hand, for $U(1)_Y$ we obtain
\be\label{udd-scal-1}
D_Y=-\dfrac{g_Y}{3}\left(-2|\tilde{u}_1|^2+|\tilde{d}_2|^2+|\tilde{d}_3|^2\right)\,.
\ee
 The scalar field content is the following,
\be\label{udd-scal-2}
\tilde{u}_1=\left( \begin{array}{c}
\varphi_1 \\
\varphi_2 \\
\varphi_3 \end{array} \right),\,
\tilde{d}_2=\left( \begin{array}{c}
\varphi_4 \\
\varphi_5 \\
\varphi_6 \end{array} \right),\,
\tilde{d}_3=\left( \begin{array}{c}
\varphi_7 \\
\varphi_8 \\
\varphi_9 \end{array} \right),
\ee
where each scalar component $\varphi_i$ is complex.
	By expanding the $D$-terms, we find the potential
\be
V_D\supset   \left(\frac{g_Y^2}{9}+\frac{g_s^2}{12}\right)\phi^2\chi_8^2
+\frac{g_s^ 2\phi^2}{12}\sum_{a=1}^7 \chi^2_a
\,,
\label{scalar-udd}
\ee
where the scalar fields $\chi_a$ are given in Eq.(\ref{8scalars}), see Appendix~\ref{DOS-udd}. The above potential gives masses proportional to the flat direction VEV, namely
\bea
&&M_{\chi_a}=\frac{g_s}{\sqrt{6}}\phi\,,
\label{amass}\\
&&M_{\chi_8}=\frac{\sqrt{3g^2_s+4g^2_Y}}{3\sqrt{2}}\phi\,.
\label{8mass}
\eea
The scalars which are coupled to the inflaton via Eq.(\ref{scalar-udd}) interact with the squarks, the Higgs doublets $H_u$ and $H_d$, the slepton doublets $\tilde{L}_i$ and the singlets $\tilde{e}_{i}$.
The interactions reported in Eqs.(\ref{scalar-udd1},\ref{scalar-udd2}) give the following decay width for $\chi_8$
\be\label{udd-dec-8}
\Gamma_{\chi_8}=\frac{3}{32\sqrt{2}\pi}  \frac{3g_s^4+8g_Y^4}{\sqrt{3g_s^2+4g_Y^2}}\phi
\,,
\ee 
while for each of the remaining seven scalars we get
\be\label{udd-dec}
\Gamma_{\chi_a}=\frac{9g_s^{3}\phi}{32\sqrt{6}\pi}\,,\qquad (a=1,...,7)
\ee
following from the SU(3) couplings in Eq.(\ref{scalar-udd2}). The decay width (\ref{udd-dec-8}) reduces to (\ref{udd-dec}) in the limit $g_Y\rightarrow 0$ indeed.

\subsection{Gauge interactions and decay widths}


Let us consider the couplings of the flat direction to the gauge fields, following from the kinetic terms \cite{Haber:1984rc,Martin:1997ns}
\be\label{gauge-kin}
\mathcal{L}\supset (D^\mu \tilde{u}_1)^\dagger(D_\mu \tilde{u}_1)+(D^\mu \tilde{d}_2)^\dagger(D_\mu \tilde{d}_2)+(D^\mu \tilde{d}_3)^\dagger(D_\mu \tilde{d}_3)\,.
\ee
Defining the covariant derivative as
\be\label{covder}
D_\mu\phi=\left(\partial_\mu-i\frac{g_Y}{2}y_kB_\mu - ig_s\,C_k\sum_{a=1}^8T^aA^a_\mu\right)\,,
\ee
for the \textit{udd} inflaton this becomes
\bea 
&&D_\mu \tilde{u}_1=
\left( \partial_\mu+\frac{2}{3}ig_Y B_\mu
-i\frac{g_s}{2}\sum_{a=1}^8\lambda^aA^a_\mu\right)\tilde{u}_1\,,\nonumber\\
&&D_\mu \tilde{d}_2=
\left( \partial_\mu-\frac{1}{3}ig_Y B_\mu
-i\frac{g_s}{2}\sum_{a=1}^8\lambda^aA^a_\mu\right)\tilde{d}_2\,,\nonumber\\
&&D_\mu \tilde{d}_3=
\left( \partial_\mu-\frac{1}{3}ig_Y B_\mu
-i\frac{g_s}{2}\sum_{a=1}^8\lambda^aA^a_\mu\right)\tilde{d}_3\,.\nonumber\\
\label{covariant}
\eea
Here $A^a_\mu$ are the eight gluons, and $B_\mu$ is the gauge field of $U(1)_Y$. These all obtain a mass proportional to the VEV of the inflaton: the flat direction VEV breaks the original $SU(3)_C\times U(1)_Y$ symmetry to a 
$U(1)$, which is a linear combination of $I_3$ and $U(1)_Y$. By using (\ref{covder}) and (\ref{covariant}), one finds indeed
\be
\mathcal{L}\supset \left(\frac{g_Y^{2}}{9}+\frac{g^2_s}{12}\right)\phi^2V_{\mu} V^{\mu}
+\frac{g^2_s\phi^2}{12}\sum_{a=1}^{7} A^a_\mu A^{a\mu}\,,
\label{gauge}
\ee
where the gauge field $V_\mu$ is defined by the rotation in field space
\be\label{V}
V_\mu = \frac{2g_YB_\mu+\sqrt{3}g_sA^8_\mu}{\sqrt{4g_Y^2+3g_s^2}}\,,
\ee
together with the gauge boson
\be\label{A}
A_\mu=\frac{2g_YA^8_\mu-\sqrt{3}g_sB_\mu}{\sqrt{4g_Y^2+3g_s^2}},
\ee
which stays massless and accounts for the residual $U(1)_Y$ symmetry. It follows that only $V_\mu$
can decay into the observable sector. These VEV-induced masses are exactly the same as for the scalars (and for the fermions), see Eqs.(\ref{amass}) and (\ref{8mass}), because they are supersymmetry conserving \cite{Allahverdi:2006xh}.

The gauge bosons which are coupled to the inflaton via Eq.(\ref{gauge}) interact with the right- and left- handed squarks, the Higgs doublets and the sleptons, together with their SUSY partners. The couplings of the gauge fields are given in Eqs.(\ref{GFS}, \ref{GFF}, \ref{Vdecscal}, \ref{Vdecscal2}, \ref{Vdecferm}, \ref{Vdecferm2}) (see Appendix~\ref{GI}). They provide with the decay width
\be\label{decayV}
\Gamma_{V}=\frac{3}{32\sqrt{2}\pi}  \frac{3g_s^4+8g_Y^4}{\sqrt{3g_s^2+4g_Y^2}}\phi
\,,
\ee
for $V_\mu$ and
\be\label{decayA}
\Gamma_{A^a_\mu}=\frac{9g_s^{3}\phi}{32\sqrt{6}\pi}\,,\qquad (a=1,...,7)
\ee
for each of the gluons $A^a_\mu$.


\subsection{Coupling to fermions and decay widths}

The flat direction couplings to the matter fermions $u_1$, $d_2$ and $d_3$ and to the gauginos, can be computed from the following part of the Lagrangian \cite{Haber:1984rc},
\bea
\mathcal{L}&\supset& \sqrt{2}
g_s\sum_{a=1}^8\left[ \tilde{u}_1^\dagger\tilde{g}_a^tT^a\left(i\sigma_2u_1 \right)
+ \tilde{d}_2^\dagger\tilde{g}_a^t T^a\left(i\sigma_2d_2 \right)\right.
\nonumber\\
&&\left.
+ \,\tilde{d}_3^\dagger\tilde{g}_a^tT^a\left(i\sigma_2d_3 \right)
\right]
-\rad\frac{g_Y}{3}\left[\,2\tilde{u}_1^\dagger\tilde{B}^t\left(i\sigma_2u_1 \right)\right.\nonumber\\
&&\left.
-\,\tilde{d}_2^\dagger\tilde{B}^t\left(i\sigma_2d_2 \right)-\tilde{d}_3^\dagger\tilde{B}^t\left(i\sigma_2d_3 \right)\right]+\mathrm{H.c.}
\label{fermlagudd}
\eea
In the above equation, $\tilde{g}_a$ and $\tilde{B}$ are the gauginos of $SU(3)_C$ and $U(1)_Y$ respectively. The field content of the quarks $u_1$, $d_2$ and $d_3$ is
\be\label{fermlagudd-1}
u_1=\left( \begin{array}{c}
\psi_1 \\
\psi_2 \\
\psi_3 \end{array} \right),\,
d_2=\left( \begin{array}{c}
\psi_4 \\
\psi_5 \\
\psi_6 \end{array} \right),\,
d_3=\left( \begin{array}{c}
\psi_7 \\
\psi_8 \\
\psi_9 \end{array} \right)\,.
\ee
In four components, Eq.\ref{fermlagudd}) can be rewritten as
\be
\mathcal{L}\supset
\frac{g_s\phi}{\sqrt{6}}\sum_{a=1}^7\bar{\Psi}_a\Psi_a
+\frac{\sqrt{3g_s^2+4g_Y^2}}{3\sqrt{2}}\phi\bar{\Psi}_V\Psi_V
\,,
\label{uddfermions}
\ee
where the 4-component spinors, $\Psi_1,...,\Psi_V$ are given by Eq.(\ref{4C8}), reported in Appendix~\ref{4CMS}.
	The decay rate of $\Psi_V$ into the squarks and quarks, Higgses and Higgsinos, leptons and sleptons is obtained from Eq.(\ref{udd-fermi-lag}) as
\be\label{ferm-decV}
\Gamma_{\Psi_V}=\frac{3}{32\sqrt{2}\pi}  \frac{3g_s^4+8g_Y^4}{\sqrt{3g_s^2+4g_Y^2}}\phi\,.
\ee
The decay width of each of the seven fermions $\Psi_a$, which interact with the right- and left-handed (s)quarks, is instead the following:
\be\label{ferm-dec}
\Gamma_{\Psi_a}=\frac{9g_s^{ 3}\phi}{32\sqrt{6}\pi}\,.
\ee
The above are consistent with the decay rates of the other fields in the SUSY multiplet that is coupled to the inflaton. As we discuss in the next Section, they will be crucial for the study of reheating after MSSM inflation.

\section{Reheating and thermalization}\label{sec:reheating-after-mssm-inflation}

Here we discuss the energy transfer from the inflaton to the MSSM matter sector via the decay of the supersymmetric multiplet to which it is coupled, and how the Universe accordingly thermalizes. A few more details about reheating in MSSM inflation are given in Ref.\cite{Allahverdi:2011aj}, where it was shown that tachyonic preheating is irrelevant, since it ends before becoming competitive with instant preheating.

As already discussed, at the end of inflation when $H(t)\sim m_\phi$ the \textit{udd} inflaton starts oscillating about the minimum of its potential, and generates bursts of coupled fields that acquire a VEV-dependent mass \cite{Felder:1998vq}, which is time-dependent due to the inflaton oscillations.

	Let us focus on the eight $\chi_a$ scalars, see Eq.(\ref{scalar-udd}), since for the gauge bosons $A_\mu^a$ and the fermions $\Psi_a$ the results will be qualitatively and quantitatively similar\footnote{
As explained in \cite{Lozanov:2016pac}, for any oscillating scalar condensate the transverse and longitudinal modes of the gauge fields coupled to the scalar can be resonantly amplified at similar, but not identical, rates. The effective frequency (\ref{eigenmodes}) thus holds for the eight $\chi$ fields and for the transverse gauge modes, while for the longitudinal gauge modes it is slightly different. Nevertheless, the amplification rates of the two types of modes are similar, and do not affect our results qualitatively \cite{Lozanov:2016pac}.}.

	 The Fourier eigenmodes of the $\chi_a$ quanta produced at each crossing have the energy
\bea \label{eigenmodes}
\omega_k &\!=\!& \sqrt{k^2 + m_{\chi_a}^2 + {g^2_s {\phi (t)}^2/6}} \nonumber \\
&\!=\!& \sqrt{k^2 + m_{\chi_a}^2 + 4 m_\phi^2 \tau^2 q_a}\,,
\eea
where $\phi(t)$ is the instantaneous VEV of the inflaton and $m_{\chi_a}$ the bare (time-independent) mass of the $\chi_a$ field. The broad resonance parameter is written as
\begin{equation}\label{q}
q_a \equiv\frac{g^2_s \dot\phi_0^2}{24m_\phi^4}\gg 1\,,
\end{equation}
and $\tau=m_\phi t$ is the time after the inflaton zero-crossing.
When the adiabaticity condition is violated, ${\dot \omega}_{k} \gsim \omega^2_{k}$, the modes with $k \lsim k_{\rm max}$ are excited each time the inflaton crosses the origin, with
\be \label{momentum}
k^2_{\rm max} \simeq \frac{1}{\sqrt6} g_s \dot \phi_0 = 2m_\phi^2\,\sqrt{q_a}\,,
\ee
and ${\dot \phi}_0$ is the velocity of the inflaton at zero crossing. The growth of the occupation number of mode $k$ for the first zero-crossing holds as
\bea \nonumber
n_{k,\chi_a} &\!=\!& \exp\!\left[{-\frac{\pi\sqrt6(k^2+m_{\chi_a}^2)}{g_s \dot\phi_0}}\!\right] \\ \label{nk}
&\!=\!& \exp\!\left[{-\frac{\pi(k^2+m_{\chi_a}^2)}{2m_\phi^2\sqrt{q_a}}}\!\right] < 1\,.
\eea
This gives the total number density of particles produced immediately after the zero-crossing as
\bea \nonumber
n_{\chi_a} \!&=&\! \int_{0}^\infty\frac{d^3 k}{(2 \pi)^3} \exp\left[-\frac{\pi(k^2+m_{\chi_a}^2)}{2m_\phi^2\sqrt{q_a}}\right] \\
\!&=&\!  \frac{m_\phi^3}{2\sqrt2 \pi^3} q^{3/4}_a\,\exp\left({-\frac{\pi m_{\chi_a}^2}{2m_\phi^2\sqrt{q_a}}}\right) \,, \label{ndensity}
\eea
before the $\chi_a$ particles decay perturbatively into the lighter degrees of freedom.

	Immediately after adiabaticity is restored, which happens at
\be
t>t_{*,a}=\sqrt{\frac{\sqrt6}{g_s\dot\phi_0}} \hspace{2mm}\Rightarrow\hspace{2mm} \tau_a > \tau_{*,a} = \frac{1}{\sqrt2}\,q^{-1/4}_a\, ,
\ee
instant preheating~\cite{Felder:1998vq} occurs: $\chi_a$ particles will decay into those that have no gauge coupling to the inflaton. In the case of the $u_1 d_2 d_3$ inflaton, these are the quarks $u_2$, $u_3$, $d_1$, the Higgses, the leptons and the according superpartners.

\begin{figure}[t]
		\centering
		\includegraphics[width=0.4\textwidth]{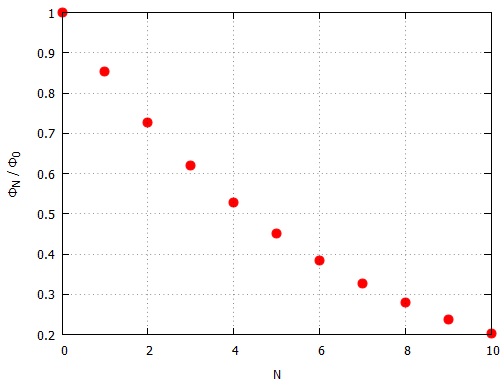}
		\caption{Evolution of the ratio $\phi_N/\phi_0$ with the number of origin crossings $N$, where $\phi_0$ is the initial oscillation amplitude.}
		\label{fig:inflatonvev}
	\end{figure}

In the case of the $\chi_8$, $V_\mu$ and $\Psi_V$ particles, since their inflaton-induced mass is
\be
M_{\chi_8}=M_{V_\mu}=M_{\Psi_V}=\frac{\sqrt{4g_Y^2+3g_s^2}}{3\rad}\phi\,,
\ee
the energy of the Fourier eigenmodes is
\begin{eqnarray} \label{eigenmodes8}
\omega_k &\!=\!& \sqrt{k^2 + m_{\chi_8}^2 + (4g^2_Y+3g_s^2) \phi (t)^2/18} \nonumber \\
&\!=\!& \sqrt{k^2 + m_{\chi_a}^2 + 4 m_\phi^2 \tau^2 q_8}\,,
\end{eqnarray}
with the broad resonance parameter
\be
q_8=\frac{4g^2_Y+3g_s^2}{72m_\phi^4}\dot{\phi}_0^2\,.
\ee
The growth of the occupation number is accordingly
\be
n_{k,\chi_8}=\exp\!\left[{-\frac{\pi(k^2+m_{\chi_8}^2)}{2m_\phi^2\sqrt{q_8}}}\!\right] < 1\,,
\ee
that returns the total number density
\be
n_{\chi_8}=\frac{m_\phi^3}{2\sqrt2 \pi^3} q^{3/4}_8\,\exp\left({-\frac{\pi m_{\chi_8}^2}{2m_\phi^2\sqrt{q_8}}}\right) \, . \label{ndensity8}
\ee
It follows immediately that the multiplet corresponding to the residual $U(1)$ symmetry is slightly heavier,
\be
\frac{M_{\chi_8}}{M_{\chi_a}}=\sqrt{1+\frac{4g_Y^2}{3g_s^2}}\sim 1.36\,,
\ee
since $g_Y\sim 0.6$ and $g_s\sim 0.75$ at the energy scale $\mathcal{O}(10^{15}\,\mathrm{GeV})$ consistent with the inflaton oscillations, see also Fig.\ref{fig:inflatonvev}. Accordingly, the decay of the gauge multiplet ($\chi_8,V_\mu,\Psi_V$) begins a bit earlier,
\be
\frac{\tau_8}{\tau_a}=\left( \frac{q_8}{q_a} \right)^{-1/4}=\left( 1+\frac{4g_Y^2}{3g_s^2} \right)^{-1/4}\sim 0.85\,.
\ee

After crossing the origin and producing bursts of quanta $\chi_a$ etc., the inflaton rolls back the potential to increase the instantaneous mass induced, together with the decay width $\Gamma_{\chi_a}$. Assuming $t_{\rm dec,a} \ll m^{-1}_{\phi}$, at the time of decay $\phi(t_{\rm dec,a}) \simeq \dot \phi_0\, t_{\rm dec,a}$, where $t_{\rm dec,a} \sim \Gamma^{-1}_{\chi_a}$
is the time between the zero crossing and $\chi_a$ decay.

Recalling (\ref{udd-dec}), namely $\Gamma_{\chi_a}=9g_s^3/(32\sqrt{6}\pi)\phi$, we find
\be \label{dectime}
1 \gg \tau_{\rm dec,a} = m_\phi t_{\rm dec,a} = \frac{4\sqrt{\pi}}{3g_s}\,q^{-1/4}_a \gg \tau_{*,a} \, .
\ee
Since the inflaton mass is $m_\phi \sim 1000\,\G$, and recall that its VEV at the beginning of oscillations is ${\hat \phi} \sim \phi_0 \simeq 10^{15}\,\G$, then $t_{\rm dec,a} \sim 6 \times 10^{-6} m^{-1}_\phi$. This guarantees the prompt decay of the $\chi_a$'s after they are produced.
	The energy density in $\chi_a$ particles soon after zero crossing is integrated as 
\bea \nonumber
\rho_{\chi_a}(\tau) &\!=\!& \int_0^\infty \frac{k^2\,dk}{2\pi^2} n_{k,\chi_a}\,\omega_k(\tau)\\
&\!=\!&  \frac{q_a\,m_\phi^4}{\pi^4}\,A_a e^{A_a} K_1(A_a)\,\exp\left({-\frac{\pi m_{\chi_a}^2}{2m_\phi^2\sqrt{q_a}}}\right) \, , \nonumber \\
\, \label{edensity}
\eea
where $K_1(z)$ is the modified Bessel function of the second kind and
\begin{eqnarray}
A_a(\tau) \equiv \frac{\pi m_{\chi_a}^2}{4 \sqrt{q_a} m_\phi^2}+ \pi \sqrt{q_a} \, \tau^2 \simeq \pi\sqrt{q_a} \, \tau^2 \, .
\end{eqnarray}
%
%
Now, $\rho_{\chi_a}$ decays with the decay rate $\Gamma_{\chi_a}$,
\bea \nonumber
&&\rho_{\chi_a}(\tau) = \rho_{\chi_a}\,{\rm exp}\!\left[{-\int_0^\tau \Gamma_{\chi_a} dt}\!\right] \\
&& =
\frac{q_a\,m_\phi^4}{\pi^4}\,A_a e^{A_a} K_1(A_a)\, e^{-A_a/2A_{\rm dec,a}} \,
\exp\!\left[{-\frac{\pi m_{\chi_a}^2}{2m_\phi^2\sqrt{q_a}}}\!\right] \, \nonumber \\
&& \,
\eea
where $A_{\rm dec, a} \equiv A_a(\tau_{\rm dec})= (16\pi^2/9g_s^2) \simeq 31.19$. When integrated beyond the time of decay, this gives
\bea \label{densitydec}
\bar\rho_{\chi_a} \!\simeq\! 8.26 \times \frac{q_a\,m_\phi^4}{\pi^4}\,\exp\!\left[{-\frac{\pi m_{\chi_a}^2}{2m_\phi^2\sqrt{q_a}}}\!\right] \, .
\eea
One can thus compute the energy ratio transferred from the inflaton to $\chi_a$'s and then into relativistic squarks at every inflaton zero-crossing:
\be \label{transfer}
\frac{\bar\rho_{\chi_a}}{\rho_\phi} \sim 0.0071\,g_s^2\,\exp\!\left[{-\frac{\pi m_{\chi_a}^2}{2m_\phi^2\sqrt{q_a}}}\!\right]\,,
\qquad \mathrm{udd}
\ee
since $\rho_\phi=\dot\phi_0^2/2$ is the inflaton energy at each zero-crossing.
%

All of the above, including Eqs.(\ref{dectime}), (\ref{densitydec}) and (\ref{transfer}) also hold for the $A^a_\mu$ gauge bosons and $\Psi_a$ fermions (a=1,...,8), which all have the same mass and decay width as $\chi_a$. Moreover, single-crossing occupation numbers in instant preheating are insensitive to the spin of the field \cite{Allahverdi:2010xz}.

	Regarding the $\chi_8$ scalar, $V_\mu$ gauge field and $\Psi_V$ fermion instead, by using (\ref{udd-dec-8}) the decay time $t_{dec}\sim \Gamma_{\chi_8}^{-1}$ gives 
\be \label{dectime8}
\tau_{\rm dec,8} = m_\phi t_{\rm dec,8} \sim \frac{4\sqrt{\pi}}{3}\sqrt{\frac{3g_s^2+4g_Y^2}{3g_s^4+8g_Y^4}}\,q^{-1/4}_8 \gg \tau_{*,y}
\ee
which corresponds to
\be
A_{8}(\tau_{\rm dec,8})=A_{\rm dec,8}=\frac{16\pi^2}{9}\frac{3g_s^2+4g_Y^2}{3g_s^4+8g_Y^4} \simeq 27.42.
\ee
The total energy density accordingly holds as
\be
\bar\rho_{\chi_8} \simeq 8.91 \times \frac{q_8\,m_\phi^4}{\pi^4}\,\exp\!\left({-\frac{\pi m_{\chi_8}^2}{2m_\phi^2\sqrt{q_8}}}\!\right) ,
\label{densitydec8}
\ee 
which implies
\be \label{transfer8}
\frac{\bar\rho_{\chi_8}}{\rho_\phi} \sim 0.0025 \, (4g^2_Y+3g_s^2)\,\exp\!\left({-\frac{\pi m_{\chi_8}^2}{2m_\phi^2\sqrt{q_8}}}\!\right)\,.
\ee
It is then evident that each of the degrees of freedom of the SUSY multiplet ($\chi_8$, $V_\mu$ and $\Psi_V$) transfers more inflaton energy than in the multiplet ($\chi_a$, $A^a_\mu$ and $\Psi_a$).
\begin{figure}[t]
	\centering
	\includegraphics[width=0.4\textwidth]{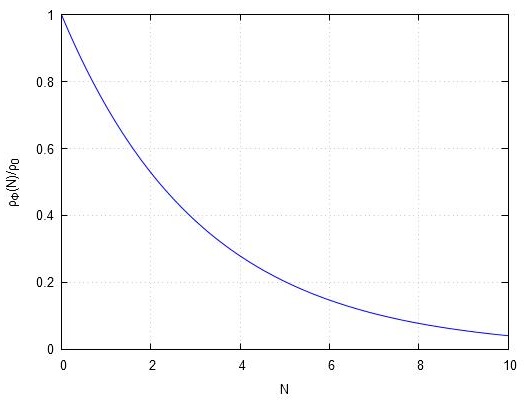}
	\caption{Evolution of the inflaton energy density with the number of origin crossings $N$, as a ratio $\rho_\phi^{(N)}/\rho_0$, where $\rho_0$ is the inflaton energy density when it starts oscillating.}
	\label{fig:energydensity}
\end{figure}

The above results can now be used to give the rate of energy transfer via instant preheating. One must add the contribution of all the degrees of freedom, namely 8 from the scalars, 24 from the 8 massive gauge bosons\footnote{The field $A_\mu$ defined in (\ref{A}), which accounts for the residual $U(1)_Y$ symmetry, is massless and stable.} and 32 from the fermions. Using Eqs.(\ref{transfer}) and (\ref{transfer8}), we can write
\be\label{rhorelvsrho}
\frac{\rho_{rel}}{\rho_\phi}=56\frac{\bar\rho_{\chi_a}}{\rho_\phi}+8\frac{\bar\rho_{\chi_8}}{\rho_\phi}\sim 28.3\%\,,
\ee
at each zero-crossing, since $g_s\sim 0.75$ and $g_Y\sim 0.6$. The process is therefore extremely efficient in transferring energy from the inflaton to the relativistic species: one can easily show that
\be\label{endens}
\rho_\phi^{(N)}=(1-0.283)^N\rho_0=(0.727)^N\rho_0\,,
\ee
where $\rho_\phi^{(N)}$ is the inflaton energy density after $N$ crossings or $N/2$ oscillations. The above means that only 4\% of the energy density at the beginning of oscillations $\rho_0$ remains in the flat direction after $N=\mathcal{O}(10)$ origin crossings, or only 5 oscillations, as pictured in Fig.\ref{fig:energydensity}.
 Notice that, by virtue of the negative exponential in (\ref{transfer}) and (\ref{transfer8}), the amplitude of oscillations is non influential.
	Compared to the analogous result for the \textit{LLe} inflaton \cite{Allahverdi:2011aj}, i.e. $\rho_{rel}/\rho_\phi\sim 10.6\%$, we see that the \textit{udd} loses energy way more efficiently.

Regarding the decreasing of the inflaton VEV with the ongoing oscillations, we can make a crude estimate neglecting the redshift effect due to the expansion of the Universe. Using conservation of energy to write the inflaton velocity when crossing the origin as $\dot{\phi}_0=\sqrt{2V(\phi)}$, an evolution equation for the inflaton VEV $\phi$ (or the oscillation amplitude) after $N$ crossings can be obtained by using $\rho_\phi=\dot{\phi}_0/2\approx \phi^2m_\phi^2/2$. Substituting into (\ref{endens}) returns indeed a very simple relation\footnote{One can show that the term $(1/2) m^2_\phi \phi^2$ is actually dominant in $V(\phi)$ even if $\phi=\mathcal{O}(10^{15}\,\G)$. Therefore we do not need $\phi$ to decrease after a certain number of oscillations to assume $\rho\approx (1/2) m^2_\phi \phi^2$ and neglect the higher powers in $\phi$.},
\be\label{fieldevolution}
\phi_N=(0.727)^{N/2}\phi_0\,,
\ee
which is plotted in Fig.\ref{fig:inflatonvev}.
In other words, after only 5 oscillations the inflaton energy density is already reduced by 96\%, and its VEV by nearly 81\%. Nevertheless, in this case $\phi_{10}=0.203\times 10^{15}\,\G$, therefore it is still very large. This is critical for gravitino production, as we show in Section \ref{sec:non-thermal-gravitino-production-in-mssm-inflation}.


	\subsection{Thermalization}
	
	In this section we address briefly the thermalization of the particles that are produced via instant preheating. This is a process that leads to thermal equilibrium of the MSSM degrees of freedom, reaching a uniform reheating temperature of the Universe.

	The main question is to determine how long it takes to reach a thermal distribution to the MSSM particle spectrum that is produced during the inflaton oscillations, as described in Sections \ref{sec:couplings} and \ref{sec:reheating-after-mssm-inflation}.
	
	The number density and energy density of the relativistic (s)leptons that are produced after the first zero-crossing of the flat direction are given respectively by
	\bea
	n_{rel}&=&56n_{\chi_a}+8n_{\chi_8}\,,\\
	\rho_{rel}&=&56\bar{\rho}_{\chi_a}+8\bar{\rho}_{\chi_8}\,,
	\eea
	with $n_{\chi_a}$, $n_{\chi_8}$, $\bar{\rho}_{\chi_a}$ and $\bar{\rho}_{\chi_8}$ given by (\ref{ndensity}), (\ref{ndensity8}), (\ref{densitydec}) and (\ref{densitydec8}). This gives
	\be\label{rhovsn}
	\rho_{rel}^{1/4}\approx 1.66\,n_{rel}^{1/3}\,.
	\ee
When these relativistic (s)leptons reach thermal equilibrium, it can be shown that their number density $n_{th}$ and energy density $\rho_{th}$ are related by $\rho_{th}^{1/4}\approx n_{th}^{1/3}$.
Compared to (\ref{rhovsn}), this means that a thermal distribution can be established only if the comoving number density of (s)leptons increases slightly. Davidson and Sarkar \cite{Davidson:2000er} showed that
this can be achieved through $2\rightarrow3$ scatterings. 

Since in this type of models we expect thermal equilibrium to be reached at around $T\approx (10^8-10^9)\,\G$, much larger than the electroweak (EW) scale, the dominant processes will be weak scatterings with exchange of massless $W$ and $Z$ bosons. Scattering of (s)quarks with gluon exchange in the t-channel is indeed initially suppressed by the inflaton VEV, which breaks the $SU(3)_C$ symmetry and gives mass to the gluons.

The according thermalization rate is written as \cite{Davidson:2000er}
\be
\Gamma_{th}^{W}\sim \alpha_W \dfrac{n_{rel}}{\sqrt{\rho_{rel}}}\,,
\ee	
	with the $SU(2)_W$ fine structure constant $\alpha_W=g_W^2/4\pi$. By substituting Eq.(\ref{rhovsn}) and (\ref{rhorelvsrho}) into the above, we find that $\Gamma_{th}^{W}\gg m_\phi$. Since the inflaton mass $m_\phi$ corresponds to the frequency of oscillations, the weak particles form a thermal bath much earlier than the next zero-crossing.
	
	The temperature of this thermal bath is initially $T=0.56\sqrt{m_\phi\phi_0}$, and increases at each zero-crossing, since 28\% of the inflaton energy density is transferred to the (s)leptons via instant preheating. Also the SUSY doublet $H_u$ and $\tilde{H}_u$ gets into thermal equilibrium with the (s)leptons because of the large Yukawa coupling of the top quark \cite{Allahverdi:2011aj}. In any case, the thermal bath does not contain the (s)quarks as their interactions are suppressed because of the broken $SU(3)_C$ symmetry, as discussed.
	
	Full thermal equilibrium indeed occurs only when all the gauge interactions have become efficient. For this to happen, the inflaton VEV must decrease enough with the oscillations so that the masses it induces on the decay products of the inflaton, see (\ref{amass}) and (\ref{8mass}), are smaller than the average kinetic energy of the particles in thermal equilibrium at all times \cite{Allahverdi:2005mz}. In other words,
	\be
\dfrac{g_s}{\sqrt{6}}\hat{\phi}\,,\, \dfrac{\sqrt{3g_s^2+4g_Y^2}}{3\rad}\hat{\phi}\lesssim 3T\,,	
	\ee
	where $\hat{\phi}$ is the oscillation amplitude. Using (\ref{rhorelvsrho}), $\rho_\phi\approx m_\phi^2\hat{\phi}^2/2$ and $\rho_{rel}=\rho_0=\pi^2/30g_* T^4$, we find that this happens after about $\mathcal{O}(40)$ oscillations. The Universe thus thermalizes very quickly, even faster for the \textit{udd} inflaton compared to other flat directions: for the \textit{LLe}  we found $\mathcal{O}(100)$ oscillations \cite{Allahverdi:2011aj}.

	Now all the MSSM degrees of freedom are kinematically available to the thermal bath, and all the gauge interactions are efficient, including $2\rightarrow2$ scatterings of colored particles. These will bring also this sector into thermal equilibrium and destroy rapidly the remaining inflaton condensate \cite{Davidson:2000er,Allahverdi:2005mz,Allahverdi:2011aj}. Accordingly, full thermalization and therefore the end of reheating will be reached after $\mathcal{O}(40)$ oscillations. This happens within a single Hubble time since $H_{\mathrm{inf}}\sim10^{-3}m_\phi$: all the MSSM degrees of freedom thermalize with a reheating temperature (the temperature when they first dominate the energy
	density of the Universe) \cite{Davidson:2000er},
	\be\label{Treh}
	T_{reh}=\left(\frac{30}{\pi^2g_\ast}\right)^{1/4}\rho_0^{1/4}\approx 2\times 10^8\,\mathrm{GeV}\,.
	\ee
	This reheating temperature is high enough to allow e.g. several mechanisms of baryogenesis \cite{Riotto:1999yt,Trodden:1998ym,Buchmuller:2012tv}.

At this point, one might wonder whether the reheating process could be even more complicated, and if there are other competing mechanisms that would affect thermalization. Q-balls \cite{Affleck:1984fy,Kasuya:1999wu,Kasuya:2000sc,Kasuya:2000wx} for example, can be generated via the fragmentation of the inflaton condensate, under some specific conditions occurring after the beginning of oscillations \cite{Enqvist:2000gq,Kusenko:1997vi,Enqvist:1998xd,Enqvist:1998ds,Enqvist:1997si,Enqvist:1998en,Doddato:2012ja}. Thus in principle they might decrease the inflaton lifetime, altering the reheating mechanism discussed above.
%
%
%
%

In our case however, an eventual Q-ball production will be suppressed by the explosive preheating discussed in this section \cite{Allahverdi:2006xh,Allahverdi:2007zz,Allahverdi:2008pf}, as it will happen late, after  baryon number creation \cite{Enqvist:2003gh,Kasuya:2001hg,Enqvist:2002rj}.

In conclusion, reheating in MSSM inflation with a \textit{udd} flat direction inflaton seems to thermalize the Universe very efficiently, with a cosmologically convenient reheating temperature.

\section{Non-thermal gravitino production in MSSM inflation}\label{sec:non-thermal-gravitino-production-in-mssm-inflation}

	The gravitino, as the supersymmetric partner of the graviton in supergravity, is a weakly interacting particle which sets important constraints on cosmological parameters. If unstable, it can decay after the beginning of BBN, thus washing out the primordial elements; if R-parity is conserved, it produces a lightest supersymmetric particle (LSP) that can account for Dark Matter (DM). The gravitino itself, in a variety of supergravity models (e.g. in gauge mediation), can be the LSP and therefore the Dark Matter particle.
	
	In this section we want therefore to investigate whether the BBN and DM constraints are respected, or if there exists a gravitino problem \cite{Khlopov:1984pf,Moroi:1995fs,Kawasaki:2006hm} in MSSM inflation.

	Gravitinos can be produced at the end of inflation either thermally, through scatterings of the particles produced by decay of the inflaton\footnote{Or produced by the decay of its decay products, like in this paper.} \cite{Bolz:2000fu,Pradler:2006qh,Kawasaki:2008qe}, or non-thermally, via the decay of the fields which determine the dynamics after inflation \cite{Kawasaki:2006gs,Kawasaki:2006hm}.

In this paper we study the latter mechanism, namely the decay into gravitinos of the supersymmetric multiplet which is coupled to the inflaton, and of the direct decay of the inflaton as well.

\subsection{Perturbative decay into gravitinos}\label{sect:perturbgravdecay}

Let us focus first on the SUSY multiplet with large induced masses $\propto g\phi$.
In general, the heavier component $\tilde{X}$ of a supermultiplet couples to its lighter superpartner $X$ and a helicity-1/2 gravitino (goldstino) through the following term of the Supergravity Lagrangian \cite{Moroi:1995fs,deGouvea:1997afu,Allahverdi:2004ds,Allahverdi:2007zz},
\be\label{goldstinolagrangian}
\mathcal{L}\supset \frac{(\Delta M)^2}{\cub \mg\mpl}\tilde{X}^*\bar{\Psi}P_R X + \mathrm{h.c.}
\ee
where $P_R \equiv (1-\gamma_5 )/2$, $\mpl=2.44\times 10^{18} \mathrm{GeV}$ is the reduced Planck mass and $\Delta M=\sqrt{m_{\tilde{X}}^2-m_X^2}\gg \mg$ is the splitting of the supersymmetry breaking masses of $\tilde{X}$ and $X$. For TeV scale SUSY, $\Delta M\approx\mathcal{O}(\T)$. $m_{\tilde{X}}$ and $m_X$ are indeed hierarchically smaller than the VEV-induced masses $M_{\tilde{X}}, M_X\propto g\phi\approx 10^{14}\G$, which are supersymmetry conserving but break the $SU(3)_C$ symmetry, as discussed in Section \ref{sec:couplings}.

Using the above coupling (\ref{goldstinolagrangian}) and taking into account the multiplicities, the fields that are coupled to the inflaton decay with the rate
\be\label{gravratesusy}
\Gamma_{\mathrm{SUSY}}\approx\frac{1}{2\pi}\frac{M_{\tilde{X}}(\Delta M)^4}{\mpl^2m_{3/2}^2}
\approx
10^{-26}\frac{\phi}{\mg^2}\qquad [\G]\,.
\ee


The inflaton field decays perturbatively into a gravitino and an inflatino. Since it is composed of 3 scalar fields, from (\ref{goldstinolagrangian}) we obtain the well known result \cite{Moroi:1995fs,deGouvea:1997afu}
\bea\label{inflatondecay}
&&\Gamma_{\mathrm{udd}}=\frac{m_\phi^5}{16\pi \mpl^2m_{3/2}^2}
\left(1-\frac{m_{3/2}^2}{m_\phi^2}\right)^2
\approx\frac{m_\phi^5}{16\pi \mpl^2m_{3/2}^2}\,,
\nonumber\\
	\eea
	where the estimate holds for light gravitinos, when the goldstino modes are dominant, as discussed above. For $m_\phi\sim\order(\T)$ this can be rewritten as
\be
\Gamma_{\mathrm{udd}}
\approx
	\frac{10^{-23}}{\mg^2}\qquad [\G].
	\ee
On the other hand Eq.(\ref{gravratesusy}) gives
\be
\Gamma_{\mathrm{SUSY}}\approx
\frac{10^{-11}}{\mg^2}\qquad [\G]\,,
\ee
where we set $\phi\sim\order (10^{15}\G)$.
Inflaton decay is therefore subdominant for most of the oscillations (see Fig.\ref{fig:inflatonvev}). Since we are investigating an eventual overproduction, we can then consider only the SUSY multiplet decay. We rewrite Eq.(\ref{gravratesusy}) as follows,
\be\label{totgravrate}
\Gamma_{3/2}\approx
10^{-17}\left(\dfrac{\mg}{1\T}\right)^{-2} \left(\dfrac{\phi}{10^{15} {\mathrm  GeV}}\right) \qquad [\G]
\,.
\ee

\subsection{Cosmological bounds}\label{sec:cosmological-bounds}

We now check explicitly that there is no gravitino problem by using the gravitino decay rate computed in the previous section.
The gravitino yield value $Y$ can be written as \cite{Nakamura:2006uc,Takahashi:2007tz}
\be\label{yieldformula}
Y\equiv \frac{n_\grav}{s}=\frac{\Gamma_\grav}{\Gamma_{\mathrm tot}}\frac{3T_R}{4m_A}\,,
\ee 
using again $m_V\approx m_A$. Since the energy contained in the inflaton is transferred to the supermultiplet at each oscillation by virtue of the mechanism described in Section \ref{sec:reheating-after-mssm-inflation}, one needs to use the decay rates of the $A-$ and $V-$ type fields with the right multiplicities. In other words,
\be\label{gammatot}
\Gamma_{\mathrm tot}=21\Gamma_A+3\Gamma_V=0.0739\phi \,,
\ee
where we took into account 3 type-$V$ particles with decay width (\ref{decayV}) and 21 type-$A$ quanta (\ref{decayA}).

Gravitino cosmology is then determined by $\phi$, the inflaton VEV, which enters the decay width via the masses (\ref{amass}) and (\ref{8mass}).
By substituting Eqs.(\ref{totgravrate}) and (\ref{gammatot}) into (\ref{yieldformula}), we obtain the yield value of the gravitino (or gravitino-to-entropy ratio) as
\be\label{yield}
Y=\dfrac{n_\grav}{s}\approx
2\times 10^{-38}\left(\dfrac{\mg}{1\T}\right)^{-2} \left(\dfrac{\phi}{10^{15} {\mathrm  GeV}}\right)^{-1}\,.
\ee
To maintain the success of the BBN, the following must hold \cite{Kawasaki:2004qu,Kawasaki:2004yh,Kawasaki:2006gs}:
\be\label{bbnconstr}
\mg Y_\grav \lesssim \order(10^{-14}\div10^{-11})\,\mathrm{GeV}\,.
\ee
On the other hand, if R-parity is conserved, an LSP gravitino is a good Dark Matter candidate if its yield value is smaller than the DM abundance, namely \cite{Allahverdi:2004si}
\be\label{dmconstr}
Y_\grav\lesssim 3\times 10^{-10}\left(\dfrac{\mathrm{1GeV}}{\mg}\right)\,.
\ee
By substituting (\ref{yield}) into (\ref{bbnconstr}) and (\ref{dmconstr}),
  we obtain
	\be\label{bbnbound}
	\mg \gtrsim \left(\dfrac{\phi}{10^{15} {\mathrm  GeV}}\right)^{-1}
	\order(10^{-21}\div10^{-18})\,\G\,,
	\ee
for the BBN constraint and
	\be\label{dmbound}
	\mg \gtrsim \left(\dfrac{\phi}{10^{15} {\mathrm  GeV}}\right)^{-1}
	\order(10^{-22})\,\G\,,
	\ee
for the DM bound. At the beginning of oscillations $\phi\sim\order (10^{15}\G)$, thus both these cosmological constraints are easily satisfied, leading to successful BBN and to a dark matter abundance not exceeding observed values.

Since we found in Section \ref{sect:perturbgravdecay} that the supermultiplet decay is strongly dominant, the above implies that also inflaton decay is non problematic, consistently with the literature \cite{Allahverdi:2000fz,Allahverdi:2008pf}.
As noted in \cite{Allahverdi:2000fz}, at early stages of inflaton oscillations only the non-thermal production of helicity $\pm 3/2$ gravitinos might be problematic to Nucleosynthesis \footnote{There is however no kinematical blocking of preheating here, because the mass scale and the couplings to the inflaton are very different from those required in \cite{Allahverdi:2007zz}. For instance, kinematical blocking occurs for $m_\phi\sim10^{13}\,\G$ and $g\sim10^{-4}$.}.

We remark that the above estimates hold for a goldstino, namely for a helicity-1/2 gravitino, corresponding to $\mg\ll \Delta M$. If we assume instead $\mg\approx \Delta M\approx 1 \T$, the decay rates of each sparticle with mass $M_{\tilde{X}}$ to both helicity-1/2 and helicity-3/2 gravitinos are comparable \cite{Moroi:1995fs}, as in our general expression for the inflaton decay rate (\ref{inflatondecay}). In the case of a sparticle $\tilde{X}$ we get
\be\label{heavygravdecay}
\Gamma_{3/2}\propto M_{\tilde{X}} \left(\frac{\mg}{\mpl}\right)^2 \approx 10^{-15}\G\,,
\ee
for a TeV-order gravitino mass and $M_{\tilde{X}}\sim\order(10^{15}\G)$. It can be easily checked that (\ref{heavygravdecay}) still satisfies the cosmological constraints.

We conclude this section with a comment on thermal gravitino production. This consists of 10 different types of scattering processes with gravitinos in the final state, which occur in the thermal bath formed according to Section \ref{sec:reheating-after-mssm-inflation} at the reheating temperature $T_R\approx2\times 10^8 \,\G$, Eq.(\ref{Treh}). This mechanism becomes efficient after $\order(40)$ oscillations and the vanishing of the inflaton VEV, and can be studied via a standard treatment that was developed in \cite{Bolz:2000fu,Pradler:2006qh,Rychkov:2007uq}\footnote{These considered respectively: QCD, the full Standard Model group, additional decays and the effect of the top Yukawa coupling.}.

In principle, also the heavy gauge and gaugino quanta might participate to the scatterings described above, and this would be dangerous because they are very massive. However, it was shown in \cite{Allahverdi:2005mz,Allahverdi:2008pf} that since these particles decay almost immediately after being produced, they will be absent from the final thermal bath. In other words, in MSSM inflation we do not expect the gravitino problem to emerge from thermal production either.

Certainly, the 10 scatterings in the thermal bath will still occur, but here $T_R$ is low enough to avoid overproduction. Yet, they might give a non-negligible gravitino abundance, depending on its mass and those of the MSSM.
The precise calculation however goes beyond the scope of this paper, since here we are only preoccupied about possible overproduction problems.

Finally, gravitino production via any eventual Q-ball decay will also be subdominant, as the reheating temperature (\ref{Treh}) is larger than $10^7\G$ \cite{Shoemaker:2009kg}.

\section{Conclusions}

In this work we have investigated particle production and thermalization in the case of a flat direction inflaton of the \textit{udd} type, thus contributing to the phenomenology of MSSM inflation for the most compelling inflaton candidates.
We computed the decay rate into the MSSM spectrum of the supersymmetric multiplet that is coupled to the flat direction, and investigated the thermalization process. We also discussed the according gravitino cosmology, which was still absent to this level of detail in previous works.

 Our calculations show that the \textit{udd} inflaton determines by far the most efficient energy transfer mechanism and reheating process with the observed degrees of freedom. About $28\%$ of the inflaton energy is drained at each origin crossing, reducing to only 4\% after 5 oscillations.
The Universe will then thermalize very quickly, after only $\mathcal{O}(40)$ oscillations and within one Hubble time, with reheating temperature $T_{reh}= \mathcal{O}(10^8\,\G)$.

	We also investigated in detail the non-thermal gravitino production from the perturbative decay of both the inflaton and the fields to which it is coupled. We found that the resulting gravitino abundance is very small and satisfies both the BBN and DM constraints. Since it is well known that thermal scatterings in MSSM inflation do not induce any gravitino overproduction \cite{Allahverdi:2005mz,Allahverdi:2008pf}, we conclude that MSSM inflation is free from any gravitino problem.

Our results are therefore encouraging, and MSSM inflation with either the \textit{LLe} or \textit{udd} inflaton seems to provide a phenomenologically interesting framework in agreement with the most recent observations.

\section{Acknowledgments}

The author wishes to thank Rouzbeh Allahverdi and Anupam Mazumdar for significant discussions and collaboration at earlier stages of this work,  and the reviewers for their valuable comments. We also acknowledge the University of Lancaster UK for the kind hospitality in several occasions.

\appendix

\section{Essential expressions for the \textit{udd} inflaton}\label{sec:essential-expressions-for-the-bf-udd-inflaton}

\subsection{Scalar interactions}
\label{DOS-udd}

The scalar fields of $SU(3)_C$ which are coupled to the inflaton are defined as follows:
\bea\label{8scalars}
&& \chi_1=\frac{\varphi_{5,R}+\varphi_{7,R}}{\rad},\qquad
\chi_2=\frac{\varphi_{5,I}-\varphi_{7,I}}{\rad},\nonumber\\
&& \chi_3=\frac{\varphi_{4,R}-\varphi_{8,R}}{\rad},\qquad
\chi_4=\frac{\varphi_{1,R}+\varphi_{6,R}}{\rad},\nonumber\\
&& \chi_5=\frac{\varphi_{1,I}-\varphi_{6,I}}{\rad},\qquad\;
\chi_6=\frac{\varphi_{2,R}+\varphi_{9,R}}{\rad},\nonumber\\
&& \chi_7=\frac{\varphi_{2,I}-\varphi_{9,I}}{\rad},\;\;
\nonumber\\
&&
\chi_8=\sqrt{\frac{2}{3}}\left(\varphi_{3,R}-\half\varphi_{4,R}-\half\varphi_{8,R}\right)
\,.
\eea
The scalar $\chi_8$ decays into the right-handed squarks which are not coupled to the inflaton, to the left-handed squarks and to the Higgs doublets and sleptons via the $SU(3)_C$ and $U(1)_Y$ $D$-terms, namely through the potential
\bea\label{scalar-udd1} 
V_D&=& \half g_s^2D_C^8D_C^8+\half g_Y^2 D_Y D_Y \nonumber\\
&\supset& 
\frac{g_s^2\phi}{2\sqrt{6}}\chi^8
\left[\tilde{u}^\dagger_2\lambda^8\tilde{u}_2+\tilde{u}^\dagger_3\lambda^8\tilde{u}_3+\tilde{d}^\dagger_1\lambda^8\tilde{d}_1
\right.
\nonumber\\
&
+&
\left.\sum_{i=1}^3(\tilde{u}^\dagger_{L,i}\lambda^8\tilde{u}_{L,i}+\tilde{d}^\dagger_{L,i}\lambda^8\tilde{d}_{L,i})\right]
\nonumber\\
&
+&
\frac{g_Y^{ 2} \phi}{3\rad }\chi^8
 \left[ \frac{1}{3}(- 4\vert {\tilde u}_2\vert^2 - 4\vert {\tilde u}_3\vert^2 
+ 2\vert {\tilde d}_1\vert^2)
\right.
\nonumber\\
&
+&\left.
\frac{1}{2}\sum_{i=1}^3(\vert {\tilde u}_{i,L}\vert^2 +\vert {\tilde d}_{i,L}\vert^2)\right.\nonumber\\
&
+&\left.
\vert H_u\vert^2-\vert H_d\vert^2-\sum_{i=1}^3 \vert {\tilde L}_i\vert^2 +2\sum_{i=1}^3\vert {\tilde e}_i\vert^2
\right]
\,.
\eea
%
	The other seven scalars $\chi^a$ defined in (\ref{8scalars}), are coupled instead to the right-handed and left-handed squarks via only the $SU(3)_C$ D-terms,
\bea\label{scalar-udd2}
V_D&\supset& \sum_{a=1}^7\frac{g_s^2\phi}{2\sqrt{6}}\chi^a 
(\tilde{u}^\dagger_2\lambda^a\tilde{u}_2+\tilde{u}^\dagger_3\lambda^a\tilde{u}_3+\tilde{d}^\dagger_1\lambda^a\tilde{d}_1\nonumber\\
&+&\tilde{u}^\dagger_{L,i}\lambda^a\tilde{u}_{L,i}+\tilde{d}^\dagger_{L,i}\lambda^a\tilde{d}_{L,i}
)
\,,
\eea
where ($i=1,2,3$). An interaction term in the form $\sigma \chi \varphi^* \varphi$ between $\chi$ and a massless scalar $\varphi$ results in a decay rate $\sigma^2/16 \pi M$, where $\sigma$ is a coupling of dimension mass and $M$ is the mass of $\chi$.
%


\subsection{Gauge interactions}
\label{GI}

The gauge bosons $A_\mu^a$ ($a=1,...,7$) and $V_\mu$ which are coupled to the inflaton interact with the matter sector through the covariant derivatives (\ref{covder}). 

	The $SU(3)$ fields $A_\mu^a$ decay into the right- and left-handed squarks, namely
	\bea\label{GFS}
	& {\cal L} & \supset
\frac{i g_s}{2}\sum_{a=1}^7A^{a\mu}(\tilde{u}_2^\dagger \lambda^a\partial_\mu\tilde{u}_2
+\tilde{u}_3^\dagger \lambda^a\partial_\mu\tilde{u}_3
+\tilde{d}_1^\dagger \lambda^a\partial_\mu\tilde{d}_1
\nonumber\\
&+&
\tilde{u}_{L,i}^\dagger \lambda^a\partial_\mu\tilde{u}_{L,i}
+\tilde{d}_{L,i}^\dagger \lambda^a\partial_\mu\tilde{d}_{L,i}
) +\mathrm{H.c.}
\nonumber\\
&=&
i\frac{g_s}{2}\big[A^{1\mu}(\tilde{u}_2^{2*}\partial_\mu\tilde{u}_2^1+\tilde{u}_2^{1*}\partial_\mu\tilde{u}_2^2
+\tilde{u}_3^{2*}\partial_\mu\tilde{u}_3^1+\tilde{u}_3^{1*}\partial_\mu\tilde{u}_3^2
\nonumber\\
&+&\tilde{d}_1^{2*}\partial_\mu\tilde{d}_1^1+\tilde{d}_1^{1*}\partial_\mu\tilde{d}_1^2
+\tilde{u}_{L,i}^{2*}\partial_\mu\tilde{u}_{L,i}^1+\tilde{u}_{L,i}^{1*}\partial_\mu\tilde{u}_{L,i}^2
\nonumber\\
&+&\tilde{d}_{L,i}^{2*}\partial_\mu\tilde{d}_{L,i}^1+\tilde{d}_{L,i}^{1*}\partial_\mu\tilde{d}_{L,i}^2)\nonumber\\
&+&iA^{2\mu}(\tilde{u}_2^{2*}\partial_\mu\tilde{u}_2^1-\tilde{u}_2^{1*}\partial_\mu\tilde{u}_2^2
+\tilde{u}_3^{2*}\partial_\mu\tilde{u}_3^1-\tilde{u}_3^{1*}\partial_\mu\tilde{u}_3^2
\nonumber\\
&+&\tilde{d}_1^{2*}\partial_\mu\tilde{d}_1^1-\tilde{d}_1^{1*}\partial_\mu\tilde{d}_1^2
+\tilde{u}_{L,i}^{2*}\partial_\mu\tilde{u}_{L,i}^1-\tilde{u}_{L,i}^{1*}\partial_\mu\tilde{u}_{L,i}^2
\nonumber\\
&+&\tilde{d}_{L,i}^{2*}\partial_\mu\tilde{d}_{L,i}^1-\tilde{d}_{L,i}^{1*}\partial_\mu\tilde{d}_{L,i}^2)
\nonumber\\
&+&A^{3\mu}(\tilde{u}_2^{1*}\partial_\mu\tilde{u}_2^1-\tilde{u}_2^{2*}\partial_\mu\tilde{u}_2^2
+\tilde{u}_3^{1*}\partial_\mu\tilde{u}_3^1-\tilde{u}_3^{2*}\partial_\mu\tilde{u}_3^2
\nonumber\\
&+&\tilde{d}_1^{1*}\partial_\mu\tilde{d}_1^1-\tilde{d}_1^{2*}\partial_\mu\tilde{d}_1^2
+\tilde{u}_{L,i}^{1*}\partial_\mu\tilde{u}_{L,i}^1-\tilde{u}_{L,i}^{2*}\partial_\mu\tilde{u}_{L,i}^2
\nonumber\\
&+&\tilde{d}_{L,i}^{1*}\partial_\mu\tilde{d}_{L,i}^1-\tilde{d}_{L,i}^{2*}\partial_\mu\tilde{d}_{L,i}^2)
\nonumber\\
&+&A^{4\mu}(\tilde{u}_2^{3*}\partial_\mu\tilde{u}_2^1+\tilde{u}_2^{1*}\partial_\mu\tilde{u}_2^3
+\tilde{u}_3^{3*}\partial_\mu\tilde{u}_3^1+\tilde{u}_3^{1*}\partial_\mu\tilde{u}_3^3
\nonumber\\
&+&\tilde{d}_1^{3*}\partial_\mu\tilde{d}_1^1+\tilde{d}_1^{1*}\partial_\mu\tilde{d}_1^3
+\tilde{u}_{L,i}^{3*}\partial_\mu\tilde{u}_{L,i}^1+\tilde{u}_{L,i}^{1*}\partial_\mu\tilde{u}_{L,i}^3
\nonumber\\
&+&\tilde{d}_{L,i}^{3*}\partial_\mu\tilde{d}_{L,i}^1+\tilde{d}_{L,i}^{1*}\partial_\mu\tilde{d}_{L,i}^3)
\nonumber\\
&+&iA^{5\mu}(\tilde{u}_2^{3*}\partial_\mu\tilde{u}_2^1-\tilde{u}_2^{1*}\partial_\mu\tilde{u}_2^3
+\tilde{u}_3^{3*}\partial_\mu\tilde{u}_3^1-\tilde{u}_3^{1*}\partial_\mu\tilde{u}_3^3
\nonumber\\
&+&\tilde{d}_1^{3*}\partial_\mu\tilde{d}_1^1-\tilde{d}_1^{1*}\partial_\mu\tilde{d}_1^3
+\tilde{u}_{L,i}^{3*}\partial_\mu\tilde{u}_{L,i}^1-\tilde{u}_{L,i}^{1*}\partial_\mu\tilde{u}_{L,i}^3
\nonumber\\
&+&\tilde{d}_{L,i}^{3*}\partial_\mu\tilde{d}_{L,i}^1-\tilde{d}_{L,i}^{1*}\partial_\mu\tilde{d}_{L,i}^3)
\nonumber\\
&+&A^{6\mu}(\tilde{u}_2^{3*}\partial_\mu\tilde{u}_2^2+\tilde{u}_2^{2*}\partial_\mu\tilde{u}_2^3
+\tilde{u}_3^{3*}\partial_\mu\tilde{u}_3^2+\tilde{u}_3^{2*}\partial_\mu\tilde{u}_3^3
\nonumber\\
&+&\tilde{d}_1^{3*}\partial_\mu\tilde{d}_1^2+\tilde{d}_1^{2*}\partial_\mu\tilde{d}_1^3
+\tilde{u}_{L,i}^{3*}\partial_\mu\tilde{u}_{L,i}^2+\tilde{u}_{L,i}^{2*}\partial_\mu\tilde{u}_{L,i}^3
\nonumber\\
&+&\tilde{d}_{L,i}^{3*}\partial_\mu\tilde{d}_{L,i}^2+\tilde{d}_{L,i}^{2*}\partial_\mu\tilde{d}_{L,i}^3)
\nonumber\\
&+&iA^{7\mu}(\tilde{u}_2^{3*}\partial_\mu\tilde{u}_2^2-\tilde{u}_2^{2*}\partial_\mu\tilde{u}_2^3
+\tilde{u}_3^{3*}\partial_\mu\tilde{u}_3^2-\tilde{u}_3^{2*}\partial_\mu\tilde{u}_3^3
\nonumber\\
&+&\tilde{d}_1^{3*}\partial_\mu\tilde{d}_1^2-\tilde{d}_1^{2*}\partial_\mu\tilde{d}_1^3
+\tilde{u}_{L,i}^{3*}\partial_\mu\tilde{u}_{L,i}^2-\tilde{u}_{L,i}^{2*}\partial_\mu\tilde{u}_{L,i}^3
\nonumber\\
&+&\tilde{d}_{L,i}^{3*}\partial_\mu\tilde{d}_{L,i}^2-\tilde{d}_{L,i}^{2*}\partial_\mu\tilde{d}_{L,i}^3)
\big] +\mathrm{H.c.}
	\eea
For a massless scalar $\phi$, the decay rates of $SU(3)_C$ gauge fields into $\phi^*\phi$ pair are $g_s^2M/96\pi$, with $M$ the mass of the gauge field.

	Decay into the corresponding fermions occurs via the following interaction terms,
	\bea\label{GFF}
	 {\cal L}  &\supset&
\frac{-i g_s}{2}\sum_{a=1}^7A^{a\mu}(\bar{\Psi}_{u,2} \lambda^a\gamma_\mu P_R\Psi_{u,2}
+\bar{\Psi}_{u,3} \lambda^a\gamma_\mu P_R\Psi_{u,3}
\nonumber\\
&+&
\bar{\Psi}_{d,1} \lambda^a\gamma_\mu P_R\Psi_{d,1}
+
\bar{\Psi}_{u,i} \lambda^a\gamma_\mu P_L\Psi_{u,i}
\nonumber\\
&+& \bar{\Psi}_{d,i} \lambda^a\gamma_\mu P_L\Psi_{d,i}
)+\mathrm{H.c.}
\nonumber\\
&=&
-i\frac{g_s}{2}\big[A^{1\mu}(\bar{\Psi}_{u,2}^2\gamma_\mu P_R\Psi_{u,2}^1+\bar{\Psi}_{u,2}^1\gamma_\mu P_R\Psi_{u,2}^2
\nonumber\\
&+&\bar{\Psi}_{u,3}^2\gamma_\mu P_R\Psi_{u,3}^1+\bar{\Psi}_{u,3}^1\gamma_\mu P_R\Psi_{u,3}^2
+\bar{\Psi}_{d,1}^2\gamma_\mu P_R\Psi_{d,1}^1\nonumber\\
&+&\bar{\Psi}_{d,1}^1\gamma_\mu P_R\Psi_{d,1}^2
+\bar{\Psi}_{u,i}^2\gamma_\mu P_L\Psi_{u,i}^1+\bar{\Psi}_{u,i}^1\gamma_\mu P_L\Psi_{u,i}^2
\nonumber\\
&+&\bar{\Psi}_{d,i}^2\gamma_\mu P_L\Psi_{d,i}^1+\bar{\Psi}_{d,i}^1\gamma_\mu P_L\Psi_{d,i}^2)\nonumber\\
&+&iA^{2\mu}(\bar{\Psi}_{u,2}^2\gamma_\mu P_R\Psi_{u,2}^1-\bar{\Psi}_{u,2}^1\gamma_\mu P_R\Psi_{u,2}^2
\nonumber\\
&+&\bar{\Psi}_{u,3}^2\gamma_\mu P_R\Psi_{u,3}^1-\bar{\Psi}_{u,3}^1\gamma_\mu P_R\Psi_{u,3}^2
+\bar{\Psi}_{d,1}^2\gamma_\mu P_R\Psi_{d,1}^1
\nonumber\\
&-&\bar{\Psi}_{d,1}^1\gamma_\mu P_R\Psi_{d,1}^2
+\bar{\Psi}_{u,i}^2\gamma_\mu P_L\Psi_{u,i}^1-\bar{\Psi}_{u,i}^1\gamma_\mu P_L\Psi_{u,i}^2
\nonumber\\
&+&\bar{\Psi}_{d,i}^2\gamma_\mu P_L\Psi_{d,i}^1-\bar{\Psi}_{d,i}^1\gamma_\mu P_L\Psi_{d,i}^2)
\nonumber\\
&+&A^{3\mu}(\bar{\Psi}_{u,2}^1\gamma_\mu P_R\Psi_{u,2}^1-\bar{\Psi}_{u,2}^2\gamma_\mu P_R\Psi_{u,2}^2
\nonumber\\
&+&\bar{\Psi}_{u,3}^1\gamma_\mu P_R\Psi_{u,3}^1-\bar{\Psi}_{u,3}^2\gamma_\mu P_R\Psi_{u,3}^2
+\bar{\Psi}_{d,1}^1\gamma_\mu P_R\Psi_{d,1}^1
\nonumber\\
&-&\bar{\Psi}_{d,1}^2\gamma_\mu P_R\Psi_{d,1}^2
+\bar{\Psi}_{u,i}^1\gamma_\mu P_L\Psi_{u,i}^1-\bar{\Psi}_{u,i}^2\gamma_\mu P_L\Psi_{u,i}^2
\nonumber\\
&+&\bar{\Psi}_{d,i}^1\gamma_\mu P_L\Psi_{d,i}^1-\bar{\Psi}_{d,i}^2\gamma_\mu P_L\Psi_{d,i}^2)
\nonumber\\
&+&A^{4\mu}(\bar{\Psi}_{u,2}^3\gamma_\mu P_R\Psi_{u,2}^1+\bar{\Psi}_{u,2}^1\gamma_\mu P_R\Psi_{u,2}^3
\nonumber\\
&+&\bar{\Psi}_{u,3}^3\gamma_\mu P_R\Psi_{u,3}^1+\bar{\Psi}_{u,3}^1\gamma_\mu P_R\Psi_{u,3}^3
+\bar{\Psi}_{d,1}^3\gamma_\mu P_R\Psi_{d,1}^1\nonumber\\
&+&\bar{\Psi}_{d,1}^1\gamma_\mu P_R\Psi_{d,1}^3
+\bar{\Psi}_{u,i}^3\gamma_\mu P_L\Psi_{u,i}^1+\bar{\Psi}_{u,i}^1\gamma_\mu P_L\Psi_{u,i}^3
\nonumber\\
&+&\bar{\Psi}_{d,i}^3\gamma_\mu P_L\Psi_{d,i}^1+\bar{\Psi}_{d,i}^1\gamma_\mu P_L\Psi_{d,i}^3)
\nonumber\\
&+&iA^{5\mu}(\bar{\Psi}_{u,2}^3\gamma_\mu P_R\Psi_{u,2}^1-\bar{\Psi}_{u,2}^1\gamma_\mu P_R\Psi_{u,2}^3
\nonumber\\
&+&
\bar{\Psi}_{u,3}^3\gamma_\mu P_R\Psi_{u,3}^1-\bar{\Psi}_{u,3}^1\gamma_\mu P_R\Psi_{u,3}^3
+\bar{\Psi}_{d,1}^3\gamma_\mu P_R\Psi_{d,1}^1
\nonumber\\
&-&\bar{\Psi}_{d,1}^1\gamma_\mu P_R\Psi_{d,1}^3
+\bar{\Psi}_{u,i}^3\gamma_\mu P_L\Psi_{u,i}^1-\bar{\Psi}_{u,i}^1\gamma_\mu P_L\Psi_{u,i}^3
\nonumber\\
&+&\bar{\Psi}_{d,i}^3\gamma_\mu P_L\Psi_{d,i}^1-\bar{\Psi}_{d,i}^1\gamma_\mu P_L\Psi_{d,i}^3)
\nonumber\\
&+&A^{6\mu}(\bar{\Psi}_{u,2}^3\gamma_\mu P_R\Psi_{u,2}^2+\bar{\Psi}_{u,2}^2\gamma_\mu P_R\Psi_{u,2}^3
\nonumber\\
&+&
\bar{\Psi}_{u,3}^3\gamma_\mu P_R\Psi_{u,3}^2+\bar{\Psi}_{u,3}^2\gamma_\mu P_R\Psi_{u,3}^3
+\bar{\Psi}_{d,1}^3\gamma_\mu P_R\Psi_{d,1}^2\nonumber\\
&+&\bar{\Psi}_{d,1}^2\gamma_\mu P_R\Psi_{d,1}^3
+\bar{\Psi}_{u,i}^3\gamma_\mu P_L\Psi_{u,i}^2+\bar{\Psi}_{u,i}^2\gamma_\mu P_L\Psi_{u,i}^3
\nonumber\\
&+&\bar{\Psi}_{d,i}^3\gamma_\mu P_L\Psi_{d,i}^2+\bar{\Psi}_{d,i}^2\gamma_\mu P_L\Psi_{d,i}^3)
\nonumber\\
&+&iA^{7\mu}(\bar{\Psi}_{u,2}^3\gamma_\mu\Psi_{u,2}^2-\bar{\Psi}_{u,2}^2\gamma_\mu P_R\Psi_{u,2}^3
\nonumber\\
&+&\bar{\Psi}_{u,3}^3\gamma_\mu P_R\Psi_{u,3}^2-\bar{\Psi}_{u,3}^2\gamma_\mu P_R\Psi_{u,3}^3
+\bar{\Psi}_{d,1}^3\gamma_\mu P_R\Psi_{d,1}^2
\nonumber\\
&-&\bar{\Psi}_{d,1}^2\gamma_\mu P_R\Psi_{d,1}^3
+\bar{\Psi}_{u,i}^3\gamma_\mu P_L\Psi_{u,i}^2-\bar{\Psi}_{u,i}^2\gamma_\mu P_L\Psi_{u,i}^3
\nonumber\\
&+&\bar{\Psi}_{d,i}^3\gamma_\mu P_L\Psi_{d,i}^2-\bar{\Psi}_{d,i}^2\gamma_\mu P_L\Psi_{d,i}^3)
\big] +\mathrm{H.c.}
	\eea
Here $P_L \equiv (1 + \gamma_5)/2$ and $P_R \equiv (1 - \gamma_5)/2$ are the left- and right-chiral projectors. The up- and down-type quarks are represented by the Dirac spinors $\Psi_u$ and $\Psi_d$ as follows,
\begin{equation} \label{uddirac}
{\Psi}_{u,i} = \left(\begin{array}{cc} Q_{u,i} \\ - i \sigma_2 u^*_i \end{array}\right)~~ , ~~ {\Psi}_{d,i} = \left(\begin{array}{cc} Q_{d,i} \\ - i \sigma_2 d^*_i \end{array}\right).
\end{equation}

The gauge bosons $V_\mu$ (massive) and $A_\mu$ (massless) have both $SU(3)_C$ and $U(1)_Y$ interactions with scalars and fermions.

By using the inverse of the transformation formulas (\ref{V}) and (\ref{A}), the following holds for right- and left-handed squarks,
\begin{eqnarray}
\label{Vdecscal}
& {\cal L} & \supset 
i\sqrt{3g_s^2+4g_Y^2}V_\mu\Big[
-\frac{1}{3}(\tilde{u}_2^{3*}\partial_\mu\tilde{u}_2^3+\tilde{u}_3^{3*}\partial_\mu\tilde{u}_3^3)
\nonumber\\
&+&
\frac{1}{6}(\tilde{d}_1^{1*}\partial_\mu\tilde{d}_1^1+\tilde{d}_1^{2*}\partial_\mu\tilde{d}_1^2)\Big]
\nonumber\\
&-&
\frac{i}{6\sqrt{3g_s^2+4g_Y^2}}[ V_\mu(8g_Y^2-3g_s^2)-6\sqrt{3}g_sg_YA_\mu] 
\nonumber\\
&&
\times(\tilde{u}_2^{1*}\partial_\mu\tilde{u}_2^1+\tilde{u}_2^{2*}\partial_\mu\tilde{u}_2^2+\tilde{u}_3^{1*}\partial_\mu\tilde{u}_3^1+\tilde{u}_3^{2*}\partial_\mu\tilde{u}_3^2)
\nonumber\\
&+&
\frac{i}{3\sqrt{3g_s^2+4g_Y^2}}[ V_\mu(2g_Y^2-3g_s^2)-3\sqrt{3}g_sg_YA_\mu]
\nonumber\\
&&
\times \tilde{d}_1^{3*}\partial_\mu\tilde{d}_1^3
\nonumber\\
&+&
\frac{i}{2\sqrt{3g_s^2+4g_Y^2}}
\Big\{
\Big[ 
\left( \frac{2}{3}g_Y^2+g_s^2 \right)V_\mu+\frac{g_sg_Y}{\sqrt{3}}A_\mu
\Big]
\nonumber\\
&&
\times (\tilde{u}_{L,i}^{1*}\partial_\mu\tilde{u}_{L,i}^1+\tilde{u}_{L,i}^{2*}\partial_\mu\tilde{u}_{L,i}^2
+(u\leftrightarrow d))
\nonumber\\
&+&
\Big[ 
\left( \frac{2}{3}g_Y^2-2g_s^2 \right)V_\mu-5\frac{g_sg_Y}{\sqrt{3}}A_\mu
\Big]
\nonumber\\
&&\times(\tilde{u}_{L,i}^{3*}\partial_\mu\tilde{u}_{L,i}^3 +(u\leftrightarrow d))
\Big\}+\mathrm{H.c.}
\end{eqnarray}
while for the Higgs doublets and the sleptons we obtain
\bea\label{Vdecscal2}
& {\cal L} & \supset 
\frac{i}{2}\frac{2g_Y^2V_\mu-\sqrt{3}g_Yg_sA_\mu}{\sqrt{3g_s^2+4g_Y^2}}
\nonumber\\
&&\times
(H_u^*\partial^\mu H_u-H_d^*\partial^\mu H_d-\tilde{L}_i^*\partial^\mu\tilde{L}_i+2\tilde{e}_i^*\partial^\mu\tilde{e}_i)
\nonumber\\
&+& \mathrm{H.c.}
\eea

On the other hand, the couplings to the fermions are
\begin{eqnarray}
\label{Vdecferm}
& {\cal L} & \supset 
-i\sqrt{3g_s^2+4g_Y^2}V_\mu
\nonumber\\
&&
\Big[
-\frac{1}{3}(\bar{\Psi}_{u,2}^3\gamma_\mu P_R\Psi_{u,2}^3+\bar{\Psi}_{u,3}^3\gamma_\mu P_R\Psi_{u,3}^3)
\nonumber\\
&+&
\frac{1}{6}(\bar{\Psi}_{d,1}^1\gamma_\mu P_R\Psi_{d,1}^1+\bar{\Psi}_{d,1}^2\gamma_\mu P_R\Psi_{d,1}^2)\Big]
\nonumber\\
&+&
\frac{i}{6\sqrt{3g_s^2+4g_Y^2}}[ V_\mu(8g_Y^2-3g_s^2)-6\sqrt{3}g_sg_YA_\mu] 
\nonumber\\
&&
\times(\bar{\Psi}_{u,2}^1\gamma_\mu P_R\Psi_{u,2}^1+\bar{\Psi}_{u,2}^2\gamma_\mu P_R\Psi_{u,2}^2
\nonumber\\
&+&\bar{\Psi}_{u,3}^1\gamma_\mu P_R\Psi_{u,3}^1+\bar{\Psi}_{u,3}^2\gamma_\mu P_R\Psi_{u,3}^2)
\nonumber\\
&-&
\frac{i}{3\sqrt{3g_s^2+4g_Y^2}}[ V_\mu(2g_Y^2-3g_s^2)-3\sqrt{3}g_sg_YA_\mu]
\nonumber\\
&&
\times \bar{\Psi}_{d,1}^3\gamma_\mu P_R\Psi_{d,1}^3
\nonumber\\
&-&
\frac{i}{2\sqrt{3g_s^2+4g_Y^2}}
\Big\{
\Big[ 
\left( \frac{2}{3}g_Y^2+g_s^2 \right)V_\mu+\frac{g_sg_Y}{\sqrt{3}}A_\mu
\Big]
\nonumber\\
&&
\times (\bar{\Psi}_{u,i}^1\gamma_\mu P_L\Psi_{u,i}^1+\bar{\Psi}_{u,i}^2\gamma_\mu P_L\Psi_{u,i}^2
+(u\leftrightarrow d))
\nonumber\\
&+&
\Big[ 
\left( \frac{2}{3}g_Y^2-2g_s^2 \right)V_\mu-5\frac{g_sg_Y}{\sqrt{3}}A_\mu
\Big]
\nonumber\\
&&\times(\bar{\Psi}_{u,i}^3\gamma_\mu P_L\Psi_{u,i}^3 +(u\leftrightarrow d))
\Big\}+\mathrm{H.c.}
\end{eqnarray}
for the quarks and
\bea\label{Vdecferm2}
& {\cal L} & \supset 
-\frac{i}{2}\frac{2g_Y^2V_\mu-\sqrt{3}g_Yg_sA_\mu}{\sqrt{3g_s^2+4g_Y^2}}
\nonumber\\
&\times&
(\bar{\Psi}_{H_u}^1\gamma^\mu P_L{\Psi}_{H_u}^1+\bar{\Psi}_{H_u}^2\gamma^\mu P_L{\Psi}_{H_u}^2
\nonumber\\
&-&\bar{\Psi}_{H_d}^1\gamma^\mu P_L{\Psi}_{H_d}^1-\bar{\Psi}_{H_d}^2\gamma^\mu P_L{\Psi}_{H_d}^2
\nonumber\\
&-&
\bar{\Psi}_{l,i}^1\gamma^\mu P_L{\Psi}_{l,i}^1-\bar{\Psi}_{l,i}^2\gamma^\mu P_L{\Psi}_{l,i}^2
+2\bar{\Psi}_{l,i}^2\gamma^\mu P_R{\Psi}_{l,i}^2)
\nonumber\\
&+& \mathrm{H.c.}
\eea
for Higgsinos and leptons.
	${\Psi}^1_{l}$ and $\Psi^2_{l}$ are Dirac spinors representing respectively the neutrinos and charged leptons (the superscript on $L_i$ refers to the weak isospin component):
\begin{equation} \label{leptondirac}
{\Psi}^1_{l,i} = \left(\begin{array}{cc} L^1_i \\ 0 \end{array}\right)~~ , ~~ {\Psi}^2_{l,i} = \left(\begin{array}{cc} L^2_i \\ - i \sigma_2 e^*_i \end{array}\right).
\end{equation}
The Higgsinos are instead represented by the spinors
\begin{equation} \label{Higgdirac}
{\Psi}^1_{H_u} = \left(\begin{array}{cc} {\tilde H}^1_u \\ 0 \end{array}\right)~~ , ~~ {\Psi}^2_{H_u} =\left(\begin{array}{cc} {\tilde H}^2_u \\ 0 \end{array}\right).
\end{equation}
For a massless fermion $\psi$, the rates for the decay of $SU(3)_C$ gauge fields to ${\bar \psi} \psi$ pair are $g^2_s M/48 \pi$.

\subsection{Fermion interactions}
\label{4CMS}

The four-component Majorana spinors $\Psi_a$ of $SU(3)_C$ are defined as
\bea\label{4C8}
&&\Psi_1 =  \left(\begin{array}{c}
 \psi_1 \\  \\ -i\sigma_2\frac{\left(\tilde{g}_4+i\tilde{g}_5\right)^*}{\sqrt{2}}  \end{array}\right),\;
\Psi_2 =  \left(\begin{array}{c}
 \psi_2 \\  \\ -i\sigma_2\frac{\left(\tilde{g}_6+i\tilde{g}_7\right)^*}{\sqrt{2}}  \end{array}\right), \nonumber \\
\, \nonumber \\
&&
\Psi_3 =  \left(\begin{array}{c}
 \psi_7 \\  \\ -i\sigma_2\frac{\left(\tilde{g}_1+i\tilde{g}_2\right)^*}{\sqrt{2}}  \end{array}\right),\;
\Psi_4 =  \left(\begin{array}{c}
 \psi_9 \\  \\ -i\sigma_2\frac{\left(\tilde{g}_6-i\tilde{g}_7\right)^*}{\sqrt{2}}  \end{array}\right),\nonumber \\
\, \nonumber \\
&&
\Psi_5 =  \left(\begin{array}{c}
 \psi_5 \\  \\ -i\sigma_2\frac{\left(\tilde{g}_1-i\tilde{g}_2\right)^*}{\sqrt{2}}  \end{array}\right),\;
\Psi_6 =  \left(\begin{array}{c}
  \psi_6 \\  \\ -i\sigma_2\frac{\left(\tilde{g}_4-i\tilde{g}_5\right)^*}{\sqrt{2}}  \end{array}\right),
\nonumber \\
\, \nonumber \\
&&
\Psi_7 =  \left(\begin{array}{c}
 \frac{\psi_4-\psi_8}{\sqrt{2}} \\  \\ -i\sigma_2\tilde{g}_3^*  \end{array}\right),\,
\nonumber
\\
\eea
and
\be
\Psi_V =  \left(\begin{array}{c}
 \tilde{V} \\  \\ \sqrt{\frac{2}{3}}(-i\sigma_2)\left( \psi_3-\frac{1}{2}\psi_4-\frac{1}{2}\psi_8 \right)^*  \end{array}\right).
\ee
The fermion $\tilde{V}$ is a linear combination of the $SU(3)_C$ and $U(1)_Y$ gauginos,
\be\label{psiV}
\tilde{V} \equiv \frac{
2g_Y \tilde{B} + \sqrt{3}g_s \tilde{g}_8
}
{
\sqrt{3g_s^2+4g_Y^2}
}\,,
\ee
in analogy with the gauge field $V_\mu$, Eq.(\ref{V}).

	The couplings to scalars and fermions are the following: the spinors $\Psi_1,...,\Psi_7$ interact with the left- and right-handed squarks and quarks via the Lagrangian
\bea
\label{udd-fermi-lag}
{\cal L} & \supset &
g_s\Big[
\tilde{Q}_{u,i}^{3*}\bar{\Psi}_1P_L\Psi^1_{u,i}
+\tilde{Q}_{u,i}^{3*}\bar{\Psi}_2P_L\Psi^2_{u,i}
+\tilde{Q}_{u,i}^{2*}\bar{\Psi}_3P_L\Psi^1_{u,i}
\nonumber\\
&+&
\tilde{Q}_{u,i}^{2*}\bar{\Psi}_4P_L\Psi^3_{u,i}
+\tilde{Q}_{u,i}^{1*}\bar{\Psi}_5P_L\Psi^2_{u,i}
+\tilde{Q}_{u,i}^{1*}\bar{\Psi}_6P_L\Psi^3_{u,i}
\nonumber\\
&+&
\tilde{Q}_{d,i}^{3*}\bar{\Psi}_1P_L\Psi^1_{d,i}
+\tilde{Q}_{d,i}^{3*}\bar{\Psi}_2P_L\Psi^2_{d,i}
+\tilde{Q}_{d,i}^{2*}\bar{\Psi}_3P_L\Psi^1_{d,i}
\nonumber\\
&+&
\tilde{Q}_{d,i}^{2*}\bar{\Psi}_4P_L\Psi^3_{d,i}
+\tilde{Q}_{d,i}^{1*}\bar{\Psi}_5P_L\Psi^2_{d,i}
+\tilde{Q}_{d,i}^{1*}\bar{\Psi}_6P_L\Psi^3_{d,i}
\nonumber\\
&+&\tilde{u}_2^{3*}\bar{\Psi}_1P_R\Psi^1_{u,2}
+\tilde{u}_2^{3*}\bar{\Psi}_2P_R\Psi^2_{u,2}
+\tilde{u}_2^{2*}\bar{\Psi}_3P_R\Psi^1_{u,2}
\nonumber\\
&+&
\tilde{u}_2^{2*}\bar{\Psi}_4P_R\Psi^3_{u,2}
+\tilde{u}_2^{1*}\bar{\Psi}_5P_R\Psi^2_{u,2}
+\tilde{u}_2^{1*}\bar{\Psi}_6P_R\Psi^3_{u,2}
\nonumber\\
&+&\tilde{u}_3^{3*}\bar{\Psi}_1P_R\Psi^1_{u,3}
+\tilde{u}_3^{3*}\bar{\Psi}_2P_R\Psi^2_{u,3}
+\tilde{u}_3^{2*}\bar{\Psi}_3P_R\Psi^1_{u,3}
\nonumber\\
&+&
\tilde{u}_3^{2*}\bar{\Psi}_4P_R\Psi^3_{u,3}
+\tilde{u}_3^{1*}\bar{\Psi}_5P_R\Psi^2_{u,3}
+\tilde{u}_3^{1*}\bar{\Psi}_6P_R\Psi^3_{u,3}
\nonumber\\
&+&\tilde{d}_1^{3*}\bar{\Psi}_1P_R\Psi^1_{d,1}
+\tilde{d}_1^{3*}\bar{\Psi}_2P_R\Psi^2_{d,1}
+\tilde{d}_1^{2*}\bar{\Psi}_3P_R\Psi^1_{d,1}
\nonumber\\
&+&
\tilde{d}_1^{2*}\bar{\Psi}_4P_R\Psi^3_{d,1}
+\tilde{d}_1^{1*}\bar{\Psi}_5P_R\Psi^2_{d,1}
+\tilde{d}_1^{1*}\bar{\Psi}_6P_R\Psi^3_{d,1}
\nonumber
\\
&+&
\tilde{Q}_{u,i}^{1*}\bar{\Psi}_7P_L\Psi^1_{u,i}-\tilde{Q}_{u,i}^{2*}\bar{\Psi}_7P_L\Psi^2_{u,i}
\nonumber
\\
&+&
\tilde{Q}_{d,i}^{1*}\bar{\Psi}_7P_L\Psi^1_{d,i}-\tilde{Q}_{d,i}^{2*}\bar{\Psi}_7P_L\Psi^2_{d,i}
\nonumber
\\
&+&
\tilde{u}_2^{1*}\bar{\Psi}_7P_R\Psi^1_{u,2}-\tilde{u}_2^{2*}\bar{\Psi}_7P_R\Psi^2_{u,2}
\nonumber
\\
&+&
\tilde{u}_3^{1*}\bar{\Psi}_7P_R\Psi^1_{u,3}-\tilde{u}_3^{2*}\bar{\Psi}_7P_R\Psi^2_{u,3}
\nonumber
\\
&+&
\tilde{d}_1^{1*}\bar{\Psi}_7P_R\Psi^1_{d,1}-\tilde{d}_1^{2*}\bar{\Psi}_7P_R\Psi^2_{d,1}
\Big]
+ {\rm H.c.}
\eea
The fermion $\Psi_V$ interacts also with the Higgs bosons, Higgsinos, leptons and sleptons. This results in the interaction Lagrangian
\bea
{\cal L}  &\supset& 
\frac{1}{\sqrt{2}}
\frac{g_s^2}{\sqrt{3g_s^2+4g_Y^2}}\times
\nonumber
\\
&\Big(&
\tilde{Q}_{u,i}^{1*}\bar{\Psi}_VP_L\Psi^1_{u,i}+\tilde{Q}_{u,i}^{2*}\bar{\Psi}_VP_L\Psi^2_{u,i}-2\tilde{Q}_{u,i}^{3*}\bar{\Psi}_VP_L\Psi^3_{u,i}
\nonumber
\\
&+&\tilde{Q}_{d,i}^{1*}\bar{\Psi}_VP_L\Psi^1_{d,i}
+\tilde{Q}_{d,i}^{2*}\bar{\Psi}_VP_L\Psi^2_{d,i}-2\tilde{Q}_{d,i}^{3*}\bar{\Psi}_VP_L\Psi^3_{d,i}
\nonumber
\\
&+&
\tilde{u}_2^{1*}\bar{\Psi}_VP_R\Psi^1_{u,2}+\tilde{u}_2^{2*}\bar{\Psi}_VP_R\Psi^2_{u,2}
-2\tilde{u}_2^{3*}\bar{\Psi}_VP_R\Psi^3_{u,2}
\nonumber\\
&+&
\tilde{u}_3^{1*}\bar{\Psi}_VP_R\Psi^1_{u,3}+\tilde{u}_3^{2*}\bar{\Psi}_VP_R\Psi^2_{u,3}
-2\tilde{u}_3^{3*}\bar{\Psi}_VP_R\Psi^3_{u,3}
\nonumber\\
&+&
\tilde{d}_1^{1*}\bar{\Psi}_VP_R\Psi^1_{d,1}+\tilde{d}_1^{2*}\bar{\Psi}_VP_R\Psi^2_{d,1}
-2\tilde{d}_1^{3*}\bar{\Psi}_VP_R\Psi^3_{d,1}
\Big)
\nonumber\\
&+&\frac{\sqrt{2}}{3}\frac{g_Y^2}{\sqrt{3g_s^2+4g_Y^2}}\times
\nonumber\\
&\Big[&
\tilde{Q}_{u,i}^{1*}\bar{\Psi}_VP_L\Psi^1_{u,i}+\tilde{Q}_{u,i}^{2*}\bar{\Psi}_VP_L\Psi^2_{u,i}+\tilde{Q}_{u,i}^{3*}\bar{\Psi}_VP_L\Psi^3_{u,i}
\nonumber
\\
&+&\tilde{Q}_{d,i}^{1*}\bar{\Psi}_VP_L\Psi^1_{d,i}
+\tilde{Q}_{d,i}^{2*}\bar{\Psi}_VP_L\Psi^2_{d,i}+\tilde{Q}_{d,i}^{3*}\bar{\Psi}_VP_L\Psi^3_{d,i}
\nonumber
\\
&-&
4(\tilde{u}_2^{1*}\bar{\Psi}_VP_R\Psi^1_{u,2}+\tilde{u}_2^{2*}\bar{\Psi}_VP_R\Psi^2_{u,2}
+\tilde{u}_2^{3*}\bar{\Psi}_VP_R\Psi^3_{u,2}
\nonumber\\
&+&
\tilde{u}_3^{1*}\bar{\Psi}_VP_R\Psi^1_{u,3}+\tilde{u}_3^{2*}\bar{\Psi}_VP_R\Psi^2_{u,3}
+\tilde{u}_3^{3*}\bar{\Psi}_VP_R\Psi^3_{u,3})
\nonumber\\
&+&
2(\tilde{d}_1^{1*}\bar{\Psi}_VP_R\Psi^1_{d,1}+\tilde{d}_1^{2*}\bar{\Psi}_VP_R\Psi^2_{d,1}
+\tilde{d}_1^{3*}\bar{\Psi}_VP_R\Psi^3_{d,1})
\Big]
\nonumber\\
&+&
\frac{\sqrt{2}g_Y^2}{\sqrt{3g_s^2+4g_Y^2}}\times
\nonumber\\
&\Big(&
H_u^{1*}\bar{\Psi}_VP_L\Psi^1_{H_u}+H_u^{2*}\bar{\Psi}_VP_L\Psi^2_{H_u}
\nonumber\\
&-&
H_d^{1*}\bar{\Psi}_VP_L\Psi^1_{H_d}-H_d^{2*}\bar{\Psi}_VP_L\Psi^2_{H_d}
\nonumber\\
&+&
\tilde{L}_i^{1*}\bar{\Psi}_VP_L\Psi^1_{l,i}+\tilde{L}_i^{2*}\bar{\Psi}_VP_L\Psi^2_{l,i}
\nonumber\\
&+&2\tilde{e}_i^{*}\bar{\Psi}_VP_R\Psi^2_{l,i}
\Big)
+\mathrm{H.c.}
\eea

The above interaction terms give a decay rate in the form $g^2m_\Psi/32\pi$.




\begin{thebibliography}{99}





\bibitem{Martin:1997ns}
S.~P.~Martin,
Adv.\ Ser.\ Direct.\ High Energy Phys.\  {\bf 21} (2010) 1
[Adv.\ Ser.\ Direct.\ High Energy Phys.\  {\bf 18} (1998) 1]
doi:10.1142/9789812839657-0001, 10.1142/9789814307505-0001
[hep-ph/9709356].

\bibitem{Haber:1984rc}
H.~E.~Haber and G.~L.~Kane,
Phys.\ Rept.\  {\bf 117} (1985) 75.
doi:10.1016/0370-1573(85)90051-1

\bibitem{Dine:2003ax}
M.~Dine and A.~Kusenko,
Rev.\ Mod.\ Phys.\  {\bf 76} (2003) 1
doi:10.1103/RevModPhys.76.1
[hep-ph/0303065].

\bibitem{Gherghetta:1995dv}
T.~Gherghetta, C.~F.~Kolda and S.~P.~Martin,
Nucl.\ Phys.\ B {\bf 468} (1996) 37
doi:10.1016/0550-3213(96)00095-8
[hep-ph/9510370].

\bibitem{Enqvist:2003gh}
K.~Enqvist and A.~Mazumdar,
Phys.\ Rept.\  {\bf 380} (2003) 99
doi:10.1016/S0370-1573(03)00119-4
[hep-ph/0209244].

\bibitem{Mazumdar:2010sa}
A.~Mazumdar and J.~Rocher,
Phys.\ Rept.\  {\bf 497} (2011) 85
doi:10.1016/j.physrep.2010.08.001
[arXiv:1001.0993 [hep-ph]].


\bibitem{Allahverdi:2010xz}
R.~Allahverdi, R.~Brandenberger, F.~Y.~Cyr-Racine and A.~Mazumdar,
Ann.\ Rev.\ Nucl.\ Part.\ Sci.\  {\bf 60} (2010) 27
doi:10.1146/annurev.nucl.012809.104511
[arXiv:1001.2600 [hep-th]].


\bibitem{Allahverdi:2006iq}
R.~Allahverdi, K.~Enqvist, J.~Garcia-Bellido and A.~Mazumdar,
Phys.\ Rev.\ Lett.\  {\bf 97} (2006) 191304
doi:10.1103/PhysRevLett.97.191304
[hep-ph/0605035].


\bibitem{Allahverdi:2006we}
R.~Allahverdi, K.~Enqvist, J.~Garcia-Bellido, A.~Jokinen and A.~Mazumdar,
JCAP {\bf 0706} (2007) 019
doi:10.1088/1475-7516/2007/06/019
[hep-ph/0610134].


\bibitem{Felder:1998vq}
G.~N.~Felder, L.~Kofman and A.~D.~Linde,
Phys.\ Rev.\ D {\bf 59} (1999) 123523
doi:10.1103/PhysRevD.59.123523
[hep-ph/9812289].

\bibitem{Allahverdi:2000fz}
R.~Allahverdi, M.~Bastero-Gil and A.~Mazumdar,
Phys.\ Rev.\ D {\bf 64} (2001) 023516
doi:10.1103/PhysRevD.64.023516
[hep-ph/0012057].

\bibitem{Traschen:1990sw}
J.~H.~Traschen and R.~H.~Brandenberger,
Phys.\ Rev.\ D {\bf 42} (1990) 2491.
doi:10.1103/PhysRevD.42.2491

\bibitem{Shtanov:1994ce}
Y.~Shtanov, J.~H.~Traschen and R.~H.~Brandenberger,
Phys.\ Rev.\ D {\bf 51} (1995) 5438
doi:10.1103/PhysRevD.51.5438
[hep-ph/9407247].

\bibitem{Kofman:1994rk}
L.~Kofman, A.~D.~Linde and A.~A.~Starobinsky,
Phys.\ Rev.\ Lett.\  {\bf 73} (1994) 3195
doi:10.1103/PhysRevLett.73.3195
[hep-th/9405187].

\bibitem{Kofman:1997yn}
L.~Kofman, A.~D.~Linde and A.~A.~Starobinsky,
Phys.\ Rev.\ D {\bf 56} (1997) 3258
doi:10.1103/PhysRevD.56.3258
[hep-ph/9704452].

\bibitem{Allahverdi:2008pf}
R.~Allahverdi and A.~Mazumdar,
Phys.\ Rev.\ D {\bf 78} (2008) 043511
doi:10.1103/PhysRevD.78.043511
[arXiv:0802.4430 [hep-ph]].



\bibitem{Allahverdi:2011aj}
R.~Allahverdi, A.~Ferrantelli, J.~Garcia-Bellido and A.~Mazumdar,
Phys.\ Rev.\ D {\bf 83} (2011) 123507
doi:10.1103/PhysRevD.83.123507
[arXiv:1103.2123 [hep-ph]].




%

\bibitem{Khlopov:1984pf}
M.~Y.~Khlopov and A.~D.~Linde,
Phys.\ Lett.\  {\bf 138B} (1984) 265.
doi:10.1016/0370-2693(84)91656-3


\bibitem{Moroi:1995fs}
T.~Moroi,
hep-ph/9503210.

\bibitem{Kawasaki:2006hm}
M.~Kawasaki, F.~Takahashi and T.~T.~Yanagida,
Phys.\ Rev.\ D {\bf 74} (2006) 043519
doi:10.1103/PhysRevD.74.043519
[hep-ph/0605297].


\bibitem{Enqvist:2010vd}
K.~Enqvist, A.~Mazumdar and P.~Stephens,
JCAP {\bf 1006} (2010) 020
doi:10.1088/1475-7516/2010/06/020
[arXiv:1004.3724 [hep-ph]].


\bibitem{Dine:1995uk}
M.~Dine, L.~Randall and S.~D.~Thomas,
Phys.\ Rev.\ Lett.\  {\bf 75} (1995) 398
doi:10.1103/PhysRevLett.75.398
[hep-ph/9503303].

\bibitem{Dine:1995kz}
M.~Dine, L.~Randall and S.~D.~Thomas,
Nucl.\ Phys.\ B {\bf 458} (1996) 291
doi:10.1016/0550-3213(95)00538-2
[hep-ph/9507453].
%

\bibitem{Wang:2013qti}
L.~Wang, E.~Pukartas and A.~Mazumdar,
JCAP {\bf 1307} (2013) 019
doi:10.1088/1475-7516/2013/07/019
[arXiv:1303.5351 [hep-ph]].


\bibitem{Ade:2013zuv}
P.~A.~R.~Ade {\it et al.} [Planck Collaboration],
Astron.\ Astrophys.\  {\bf 571} (2014) A16
doi:10.1051/0004-6361/201321591
[arXiv:1303.5076 [astro-ph.CO]].

\bibitem{Allahverdi:2006xh}
R.~Allahverdi and A.~Mazumdar,
JCAP {\bf 0708} (2007) 023
doi:10.1088/1475-7516/2007/08/023
[hep-ph/0608296].



\bibitem{Lozanov:2016pac}
K.~D.~Lozanov and M.~A.~Amin,
JCAP {\bf 1606} (2016) no.06,  032
doi:10.1088/1475-7516/2016/06/032
[arXiv:1603.05663 [hep-ph]].


\bibitem{Davidson:2000er}
S.~Davidson and S.~Sarkar,
JHEP {\bf 0011} (2000) 012
doi:10.1088/1126-6708/2000/11/012
[hep-ph/0009078].

\bibitem{Allahverdi:2005mz}
R.~Allahverdi and A.~Mazumdar,
JCAP {\bf 0610} (2006) 008
doi:10.1088/1475-7516/2006/10/008
[hep-ph/0512227].

\bibitem{Riotto:1999yt}
A.~Riotto and M.~Trodden,
Ann.\ Rev.\ Nucl.\ Part.\ Sci.\  {\bf 49} (1999) 35
doi:10.1146/annurev.nucl.49.1.35
[hep-ph/9901362].

\bibitem{Trodden:1998ym}
M.~Trodden,
Rev.\ Mod.\ Phys.\  {\bf 71} (1999) 1463
doi:10.1103/RevModPhys.71.1463
[hep-ph/9803479].

\bibitem{Buchmuller:2012tv}
W.~Buchm\"uller,
Acta Phys.\ Polon.\ B {\bf 43} (2012) 2153
doi:10.5506/APhysPolB.43.2153
[arXiv:1212.3554 [hep-ph]].


\bibitem{Affleck:1984fy}
I.~Affleck and M.~Dine,
Nucl.\ Phys.\ B {\bf 249} (1985) 361.
doi:10.1016/0550-3213(85)90021-5




\bibitem{Kasuya:1999wu}
S.~Kasuya and M.~Kawasaki,
Phys.\ Rev.\ D {\bf 61} (2000) 041301
doi:10.1103/PhysRevD.61.041301
[hep-ph/9909509].

\bibitem{Kasuya:2000sc}
S.~Kasuya and M.~Kawasaki,
Phys.\ Rev.\ Lett.\  {\bf 85} (2000) 2677
doi:10.1103/PhysRevLett.85.2677
[hep-ph/0006128].

\bibitem{Kasuya:2000wx}
S.~Kasuya and M.~Kawasaki,
Phys.\ Rev.\ D {\bf 62} (2000) 023512
doi:10.1103/PhysRevD.62.023512
[hep-ph/0002285].



\bibitem{Enqvist:2000gq}
K.~Enqvist, A.~Jokinen and J.~McDonald,
Phys.\ Lett.\ B {\bf 483} (2000) 191
doi:10.1016/S0370-2693(00)00578-5
[hep-ph/0004050].



\bibitem{Kusenko:1997vi}
A.~Kusenko, M.~E.~Shaposhnikov and P.~G.~Tinyakov,
Pisma Zh.\ Eksp.\ Teor.\ Fiz.\  {\bf 67} (1998) 229
[JETP Lett.\  {\bf 67} (1998) 247]
doi:10.1134/1.567658
[hep-th/9801041].



\bibitem{Enqvist:1998xd}
K.~Enqvist and J.~McDonald,
Phys.\ Lett.\ B {\bf 440} (1998) 59
doi:10.1016/S0370-2693(98)01078-8
[hep-ph/9807269].

\bibitem{Enqvist:1998ds}
K.~Enqvist and J.~McDonald,
Phys.\ Rev.\ Lett.\  {\bf 81} (1998) 3071
doi:10.1103/PhysRevLett.81.3071
[hep-ph/9806213].

\bibitem{Enqvist:1997si}
K.~Enqvist and J.~McDonald,
Phys.\ Lett.\ B {\bf 425} (1998) 309
doi:10.1016/S0370-2693(98)00271-8
[hep-ph/9711514].

\bibitem{Enqvist:1998en}
K.~Enqvist and J.~McDonald,
Nucl.\ Phys.\ B {\bf 538} (1999) 321
doi:10.1016/S0550-3213(98)00695-6
[hep-ph/9803380].


\bibitem{Doddato:2012ja}
F.~Doddato and J.~McDonald,
JCAP {\bf 1307} (2013) 004
doi:10.1088/1475-7516/2013/07/004
[arXiv:1211.1892 [hep-ph]].

\bibitem{Allahverdi:2007zz}
R.~Allahverdi and A.~Mazumdar,
Phys.\ Rev.\ D {\bf 76} (2007) 103526
doi:10.1103/PhysRevD.76.103526
[hep-ph/0603244].

\bibitem{Kasuya:2001hg}
S.~Kasuya and M.~Kawasaki,
Phys.\ Rev.\ D {\bf 64} (2001) 123515
doi:10.1103/PhysRevD.64.123515
[hep-ph/0106119].
%




\bibitem{Enqvist:2002rj}
K.~Enqvist, S.~Kasuya and A.~Mazumdar,
Phys.\ Rev.\ Lett.\  {\bf 89} (2002) 091301
doi:10.1103/PhysRevLett.89.091301
[hep-ph/0204270].


%
%


\bibitem{Bolz:2000fu}
M.~Bolz, A.~Brandenburg and W.~Buchmuller,
Nucl.\ Phys.\ B {\bf 606} (2001) 518
Erratum: [Nucl.\ Phys.\ B {\bf 790} (2008) 336]
doi:10.1016/S0550-3213(01)00132-8, 10.1016/j.nuclphysb.2007.09.020
[hep-ph/0012052].

\bibitem{Pradler:2006qh}
J.~Pradler and F.~D.~Steffen,
Phys.\ Rev.\ D {\bf 75} (2007) 023509
doi:10.1103/PhysRevD.75.023509
[hep-ph/0608344].

\bibitem{Kawasaki:2008qe}
M.~Kawasaki, K.~Kohri, T.~Moroi and A.~Yotsuyanagi,
Phys.\ Rev.\ D {\bf 78} (2008) 065011
doi:10.1103/PhysRevD.78.065011
[arXiv:0804.3745 [hep-ph]].

\bibitem{Kawasaki:2006gs}
M.~Kawasaki, F.~Takahashi and T.~T.~Yanagida,
Phys.\ Lett.\ B {\bf 638} (2006) 8
doi:10.1016/j.physletb.2006.05.037
[hep-ph/0603265].



\bibitem{deGouvea:1997afu}
A.~de Gouvea, T.~Moroi and H.~Murayama,
Phys.\ Rev.\ D {\bf 56} (1997) 1281
doi:10.1103/PhysRevD.56.1281
[hep-ph/9701244].



\bibitem{Allahverdi:2004ds}
R.~Allahverdi and M.~Drees,
Phys.\ Rev.\ D {\bf 70} (2004) 123522
doi:10.1103/PhysRevD.70.123522
[hep-ph/0408289].




\bibitem{Nakamura:2006uc}
S.~Nakamura and M.~Yamaguchi,
Phys.\ Lett.\ B {\bf 638} (2006) 389
doi:10.1016/j.physletb.2006.05.078
[hep-ph/0602081].



\bibitem{Takahashi:2007tz}
F.~Takahashi,
Phys.\ Lett.\ B {\bf 660} (2008) 100
doi:10.1016/j.physletb.2007.12.048
[arXiv:0705.0579 [hep-ph]].



\bibitem{Kawasaki:2004qu}
M.~Kawasaki, K.~Kohri and T.~Moroi,
Phys.\ Rev.\ D {\bf 71} (2005) 083502
doi:10.1103/PhysRevD.71.083502
[astro-ph/0408426].

\bibitem{Kawasaki:2004yh}
M.~Kawasaki, K.~Kohri and T.~Moroi,
Phys.\ Lett.\ B {\bf 625} (2005) 7
doi:10.1016/j.physletb.2005.08.045
[astro-ph/0402490].


\bibitem{Allahverdi:2004si}
R.~Allahverdi, A.~Jokinen and A.~Mazumdar,
Phys.\ Rev.\ D {\bf 71} (2005) 043505
doi:10.1103/PhysRevD.71.043505
[hep-ph/0410169].




\bibitem{Rychkov:2007uq}
V.~S.~Rychkov and A.~Strumia,
Phys.\ Rev.\ D {\bf 75} (2007) 075011
doi:10.1103/PhysRevD.75.075011
[hep-ph/0701104].


\bibitem{Shoemaker:2009kg}
I.~M.~Shoemaker and A.~Kusenko,
Phys.\ Rev.\ D {\bf 80} (2009) 075021
doi:10.1103/PhysRevD.80.075021
[arXiv:0909.3334 [hep-ph]].






%
%



  
  \end{thebibliography}
\end{document}